\documentclass[fleqn,10pt]{wlscirep}
\usepackage[utf8]{inputenc}
\usepackage[T1]{fontenc}
\usepackage{amsmath}    
\usepackage{amssymb}    
\usepackage{lineno}
\usepackage{amsmath}

\usepackage{caption}
\DeclareCaptionType[fileext=ext]{Supplementary Figure}

\let\oldAA\AA
\renewcommand{\AA}{\text{\normalfont\oldAA}}

\newcommand{\nv}{N\,{\sc v}~}
\newcommand{\civ}{C\,{\sc iv}~}
\newcommand{\Heii}{He\,{\sc ii}~}

\newcommand{\oiii}{[O\,{\sc III}]~}

\def \ll {$\tt\lambda\lambda$}
\def \l {$\tt \lambda$}

\title{Detection of companion galaxies around hot dust-obscured hyper-luminous galaxy W0410-0913}

\author[1,*]{M. Ginolfi}
\author[2]{E. Piconcelli}
\author[2]{L. Zappacosta}
\author[3, 4, 5]{G. C. Jones}
\author[2]{L. Pentericci}
\author[4, 5]{R. Maiolino}
\author[6]{A. Travascio}
\author[2]{N. Menci}
\author[7]{S. Carniani}
\author[8, 9]{F. Rizzo}

\author[10]{F. Arrigoni Battaia}
\author[6]{S. Cantalupo}
\author[1]{C. De Breuck}
\author[11]{L. Graziani}
\author[12]{K. Knudsen}
\author[8, 9]{P. Laursen}
\author[1]{V. Mainieri}
\author[11]{R. Schneider}
\author[13]{F. Stanley}
\author[2]{R. Valiante}
\author[14]{A. Verhamme}

\affil[1]{European Southern Observatory, Karl-Schwarzschild-Str. 2, D-85748 Garching bei München, Germany}
\affil[2]{INAF - Osservatorio Astronomico di Roma, Via Frascati 33, I-00040 Monte Porzio Catone, Italy}
\affil[3]{Department of Physics, University of Oxford, Denys Wilkinson Building, Keble Road, Oxford, OX1 4RH, UK}
\affil[4]{Cavendish Laboratory, University of Cambridge, 19 J. J. Thomson Ave., Cambridge CB3 0HE, UK}
\affil[5]{Kavli Institute for Cosmology, University of Cambridge, Madingley Road, Cambridge CB3 0HA, UK}
\affil[6]{Department of Physics, University of Milan Bicocca, Piazza della Scienza 3, I-20126 Milano, Italy}
\affil[7]{Scuola Normale Superiore, Piazza dei Cavalieri 7, 56126, Pisa, Italy}
\affil[8]{Cosmic Dawn Center (DAWN)}
\affil[9]{Niels Bohr Institute, University of Copenhagen, Jagtvej 128, DK-2200, Copenhagen N, Denmark}
\affil[10]{Max-Planck-Institut für Astrophysik, Karl-Schwarzschild-Str 1, D-85748 Garching bei München, Germany}
\affil[11]{Dipartimento di Fisica, Sapienza, Universita di Roma, Piazzale Aldo Moro 5, I-00185 Roma, Italy}
\affil[12]{Department of Space, Earth, and Environment, Chalmers University of Technology, Onsala Space Observatory, SE-439 92 Onsala, Sweden}
\affil[13]{Sorbonne Université, UPMC Université Paris 6 \& CNRS, UMR 7095, Institut d’Astrophysique de Paris, 98b boulevard Arago,
	75014 Paris, France}
\affil[14]{Observatoire de Genéve, Université de Genève, 51 Ch. des Maillettes, 1290 Versoix, Switzerland}

 \affil[*]{corresponding author: Michele Ginolfi (michele.ginolfi@eso.org)}

\begin{abstract} 

\bf{\normalsize{
The phase transition between galaxies and quasars is often identified with the rare population of  hyper-luminous, hot dust-obscured galaxies. 
Galaxy formation models predict these systems to grow via mergers, that can deliver large amounts of gas toward their centers, induce intense bursts of star formation and feed their supermassive black holes. 
Here we report the detection of 24 galaxies emitting Lyman-$\alpha$ emission on projected physical scales of about $400 ~{\rm kpc}$ around the hyper-luminous hot dust-obscured galaxy W0410-0913, at redshift $z = 3.631$, using Very Large Telescope observations.
While this indicates that W0410-0913 evolves in a very dense environment, we do not find clear signs of mergers that could sustain its growth. 
Data suggest that if mergers occurred, as models expect, these would involve less massive satellites, with only a moderate impact on the internal interstellar medium of W0410-0913, which is sustained by a rotationally-supported fast-rotating molecular disk, as Atacama Large Millimeter Array observations suggest.
}}

\end{abstract}

\begin{document}

\flushbottom
\maketitle

\thispagestyle{empty}

\section*{Article}

\subsection*{Introduction}

Observations carried out with the Wide-field Infrared Survey Explorer (WISE) \cite{Wright2010} recently led to the discovery of a population of hyper-luminous dust-obscured quasars with bolometric luminosities of $L_{\rm bol} \gtrsim 10^{13} - 10^{14} ~ {\rm L_\odot}$ \cite{Wu2012, Eisenhardt2012}, characterized by hot dust temperatures ($T_{\rm dust} > 60$ K) and very red mid- infrared (IR) colors.
These quasars are thought to be powered by buried active galactic nuclei (AGNs), with accretion rates close to the Eddington limit \cite{Assef2015}.
There is a growing consensus that this rare population of extreme objects, often referred to as Hot Dust-Obscured Galaxies (Hot DOGs) \cite{Wu2012}, represents a unique and short-lived stage of galaxy evolution in which both heating and ejection of dusty interstellar medium (ISM) through winds powered by the central AGN are caught in the act \cite{Finnerty2020} (the so-called blowout phase \cite{Hopkins2008, Bridge2013}), leading to an evolutionary transition from dusty starburst galaxies to ultraviolet (UV)-bright unobscured quasars, and finally to giant elliptical galaxies \cite{Hopkins2008}.

Models predict that the growth of these massive star-bursting objects is driven by galaxy mergers \cite{Hopkins2006, DiMatteo2008}, which can deliver a significant amount of gas toward their centers, boosting their star formation rate (SFR) up to thousands of ${\rm M_\odot~yr^{-1}}$ ~\cite{Frey2016}, and feeding the growth of their supermassive black holes (SMBH), thereby powering the AGN activity \cite{Hopkins2006}.
However, while cumulative number counts in wide-field surveys suggest that Hot DOGs tend to lie in dense large-scale environments \cite{Fan2017, Jones2014}, direct observations of the mechanisms responsible for their rapid growth are still sparse or only restricted to scales of a few tens of kpc \cite{DiazSantos2018}.
On the other hand, recent works \cite{Bischetti2018, Hill2020, GarciaVergara2022} have demonstrated the potential of spectroscopic imaging studies across the environment of luminous quasars and protocluster galaxies, finding companion galaxies out to hundreds of kpc.
\newline 

In this work, we report rest-frame UV and rest-frame fir-infrared (FIR) spectroscopic imaging observations of the Hot DOG WISE J041010.60-0913056.2 (hereafter W0410-0913), obtained with the integral field spectrograph Multi Unit Spectroscopic Explorer (MUSE) at the Very Large Telescope (VLT), and with the Atacama Large Millimeter/Sub-millimeter Array (ALMA), respectively.
W0410-0913 was originally selected from a WISE all-sky survey\cite{Wright2010} and, with a total FIR luminosity of $L_{\rm FIR} \sim 2\times10^{14} ~ {\rm L_\odot}$ \cite{Fan2016}, and with large stellar \cite{DiazSantos2021} ($M_{\rm star}$) and molecular gas \cite{Fan2018} ($M_{\rm gas}$) masses of $~\gtrsim 10^{11} ~ {\rm M_\odot}$, is one of the brightest, most massive and gas-rich galaxies in the distant Universe \cite{Tsai2015, Fan2018}.
Extensive rest-frame FIR observations, tracing several molecular lines, have shown W0410-0913 to be located at a systemic redshift of $z = 3.631$ \cite{Stanley2021, DiazSantos2021} (see details on the redshift determination in the subsection ALMA data in Methods), about only $1.7~ {\rm Gyr}$ after the Big Bang. 
Through the combined analysis of the Lyman-$\alpha$ transition of atomic hydrogen (at a rest-frame wavelength of 1215.67 $\AA$) and the rotational transitions of carbon monoxide, these observations allow us to study both the circumgalactic environment surrounding the hyper-luminous galaxy, on scales of $\gtrsim 100 ~{\rm kpc}$, as well as the morphological and kinematic structure of its ISM, on scales of a few ${\rm kpc}$, providing information on the multi-scale processes that drive its evolution.

\subsection*{Results}

\subsubsection*{MUSE observations of the circumgalactic environment of W0410-0913}

We have obtained deep MUSE observations of W0410-0913 (exploiting a total telescope time of 2 h) as part of an observational campaign aimed at obtaining spatially resolved rest-frame UV spectroscopy observations of a Hot DOG and its halo gas, also called circumgalactic medium (CGM).
Due to its wide field of view (FOV) and high sensitivity, MUSE is a suitable instrument to explore the surrounding environment of luminous sources in the distant Universe, and it has carried out observations of both morphology and kinematics of the gas distribution around luminous distant quasars \cite{Borisova2016, ArrigoniBattaia2019} through direct imaging of the Lyman-$\alpha$ emission, which is mainly produced through recombination radiation following photo-ionization powered by bright sources \cite{Cantalupo2014}.

The spectrum of W0410-0913 observed with MUSE shows highly attenuated rest-frame UV emission.
We detect very broad (full widths at half maxima of about $\rm FWHM \sim  2800 ~ km ~ s^{-1}$) blueshifted components of \nv \l 1240 and \civ \ll 1548,1550, both offset by more than $\rm  2000 ~ km ~ s^{-1}$ with respect to the systemic redshift, likely tracing fast AGN-driven outflows \cite{DiazSantos2018, Jun2020, Finnerty2020}.
Also, we find a narrow ($\rm FWHM \sim  400 ~ km ~ s^{-1}$) Lyman-$\alpha$ line, which, similarly to the narrow component of the optical \oiii \l  5007 line \cite{Finnerty2020}, is blueshifted by about $\rm  1400 ~ km ~ s^{-1}$.
We find evidence for a small Lyman-$\alpha$ nebula associated to this line, with a maximum projected linear size of about $30 ~{\rm kpc}$.
This is at odds with recent observations of several Lyman-$\alpha$ nebulae showing sizes up to hundreds of $\rm kpc$ detected with MUSE around UV-bright quasars, with typical $L_{\rm bol}$ even lower than W0410-0913 and shorter exposure times \cite{Borisova2016, ArrigoniBattaia2019}.
Our finding is rather to be ascribed to the large amount of dust in W0410-0913, which prevents Lyman-$\alpha$ and UV photons to escape the galaxy and illuminate the CGM, suggesting that extreme physical conditions of the ISM can have a profound impact on the properties of the CGM accessible through Lyman-$\alpha$. 

The lack of extended diffuse Lyman-$\alpha$ emission dominating the field at wavelengths close to the systemic redshift of W0410-0913 makes it possible to efficiently search for Lyman-$\alpha$-emitting companion galaxies in the MUSE datacube, and thus provides well suited conditions to explore the surrounding environment of a Hot DOG on circumgalactic scales, informing on the physics that drives the rapid evolution of this rare population of objects.
\newline

\begin{figure*}[ht]
	\centering
	\includegraphics[width=0.5\linewidth]{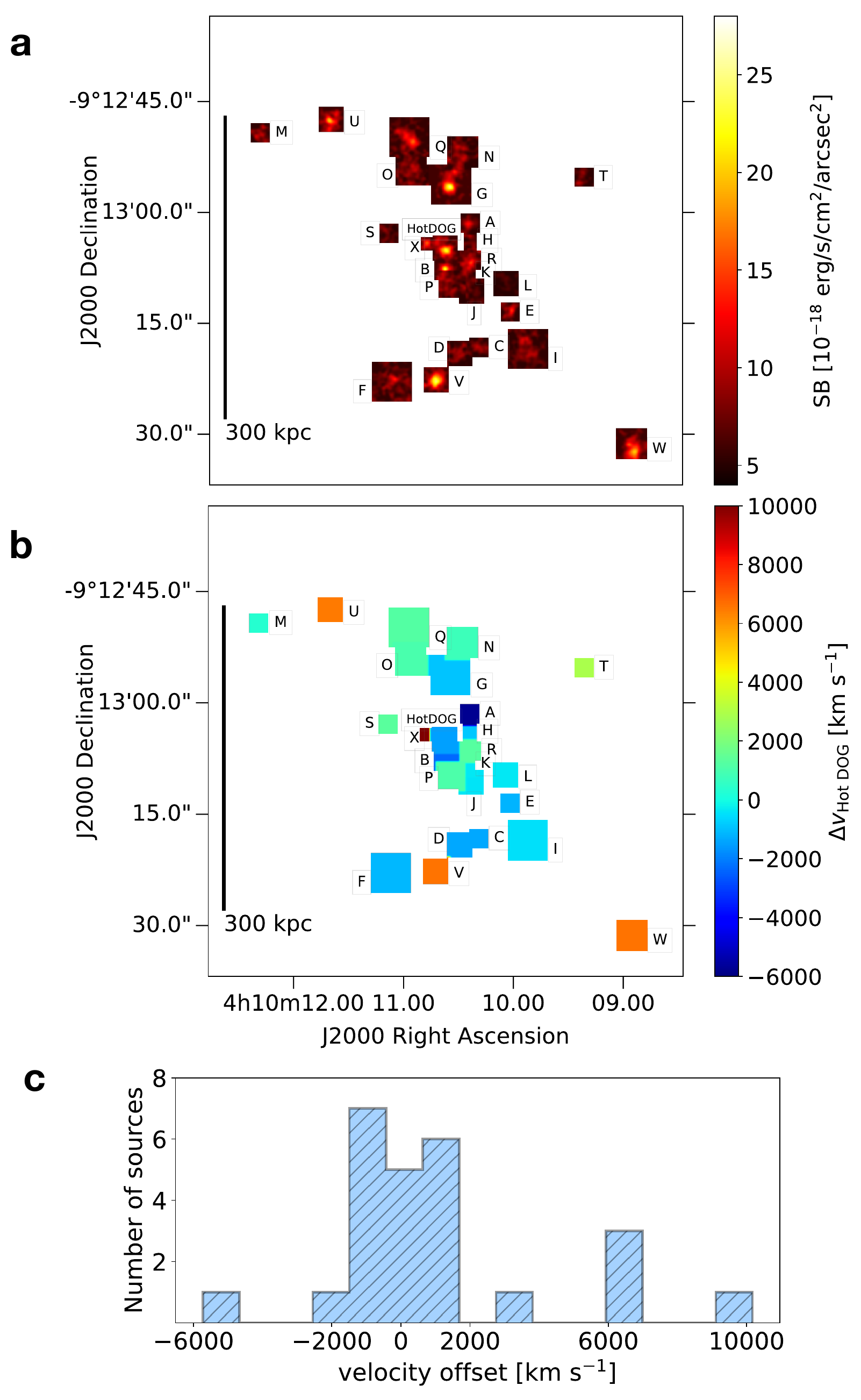}
	\caption{
		{\bf Overdensity of Lyman-$\alpha$ emitting galaxies in the circumgalactic environment around the hyper-luminous  W0410-0913, at $z=3.631$}.
		{\bf a} We show a collage of Lyman-$\alpha$ flux maps showing the projected spatial distribution of 24 Lyman-$\alpha$-emitting companion galaxies revealed by MUSE in the field of W0410-0913. 
		We applied a spatial Gaussian filtering of 1 pixel ($0.2''$) to the cutouts and we set a lower bound of ${\rm SB} = 5\times10^{-18} \rm{erg ~ s^{-1} cm^{-2} arcsec^{-2}}$ in the color-scale to improve the visualization.
		The Lyman-$\alpha$ emitters are distributed on scales of about $400~{\rm kpc}$ around the W0410-0913 (see the black line, which corresponds to $300~{\rm kpc}$).
		The size of each cutout varies between $2''$ and $4''$, and is chosen according to the extensions of the emitting sources.
		See Table 1 for the definition of IDs and a summary of the measured properties of the Lyman-$\alpha$ emitters.
		{\bf b} We color the cutouts shown in panel {\bf a} according to the offset velocities of the companion galaxies with respect to the systemic redshift of W0410-0913. The redshift of each emitter is traced by the center wavelength obtained through a Gaussian fit of the Lyman-$\alpha$ spectrum. 
			For W0410-0913, at the center, we show the velocity as traced by its Lyman-$\alpha$, which is blueshifted with respect to the systemic redshift.
		The offset velocities range between $-6.000 ~ \rm{km ~ s^{-1}}$ and $10.000 ~ \rm{km ~ s^{-1}}$.
		A histogram of their distribution is shown in panel {\bf c}.
		Most of the companion galaxies (18 out of 24) show offsets within a smaller velocity range of about $[-2.500; +1.500] ~ \rm{km ~ s^{-1}}$.
		These are more likely subject to the gravitational effect of the central massive object.
	}
	\label{fig:1}
\end{figure*}

Our MUSE observations (reaching a $2 \sigma$ surface brightness (SB) limit of $5\times10^{-19} \rm{erg ~ s^{-1} cm^{-2} arcsec^{-2}}$ in a $1 ~ \rm{arcsec^2}$ aperture and in a single spectral channel) reveal a high overdensity of 24 Lyman-$\alpha$-emitting satellite galaxies associated with W0410-0913 within the MUSE FOV, in a region of about $400$ projected physical kpc (diameter-scale) centered around the  central massive luminous object (see Figure \ref{fig:1}).
A compilation of their Lyman-$\alpha$ flux maps is shown in Figure \ref{fig:1}a, providing information on their projected spatial distribution in the field.
The adopted techniques and the analysis performed to search for these galaxies (as well as a statistical assessment of the reliability of their detections) are discussed in the subsections describing the MUSE data in Methods.

Archival images from the Hubble Space Telescope (HST) obtained with the WFC3 F160W filter (probing rest-frame emission at around $0.3 ~ {\rm \mu m}$) show that, despite the modest exposure (only $2400 ~ {\rm s}$), ten among these sources (i.e., approximately $40 \%$ of the total) have rest-frame optical counterparts detected with a signal-to-noise ratio (SNR) higher than $4 ~ \sigma$, and corresponding SFRs ranging between $\gtrsim ~ 12 - 100 ~{\rm M_\odot ~ yr^{-1}}$ (computed using standard $L_{\rm UV/optical}-{\rm SFR}$ calibrations; see subsection Measure of the SFR from HST in Methods).
Most of the remaining objects show only tentative individual detections at SNR of $2-3 ~\sigma$, but from a stacking analysis we can infer an average ${\rm SFR}$ of about  $3 ~{\rm M_\odot ~ yr^{-1}}$.

MUSE data show that the companion galaxies discovered around W0410-0913 have offset line-of-sight velocities, traced by the peak of Lyman-$\alpha$ in their spectra (see Methods, where resonant scattering effects are also discussed), in the broad range of about $[-6.000; +10.000] ~ \rm{km ~ s^{-1}}$ with respect to the systemic redshift of the central galaxy (see Figure \ref{fig:1}b and Figure \ref{fig:1}c), traced by its molecular gas. 
75 \% of them (18 out of 24) show offsets in a smaller velocity range of about $[-2.500; +1.500] ~ \rm{km ~ s^{-1}}$ (see subsection The circumgalactic environment of W0410-0913 observed with MUSE in Methods; see Figure \ref{fig:1}c), suggesting that these satellites are experiencing a strong attraction by W0410-0913.

\begin{figure*}[ht]
	\centering
	\includegraphics[width=0.8\linewidth]{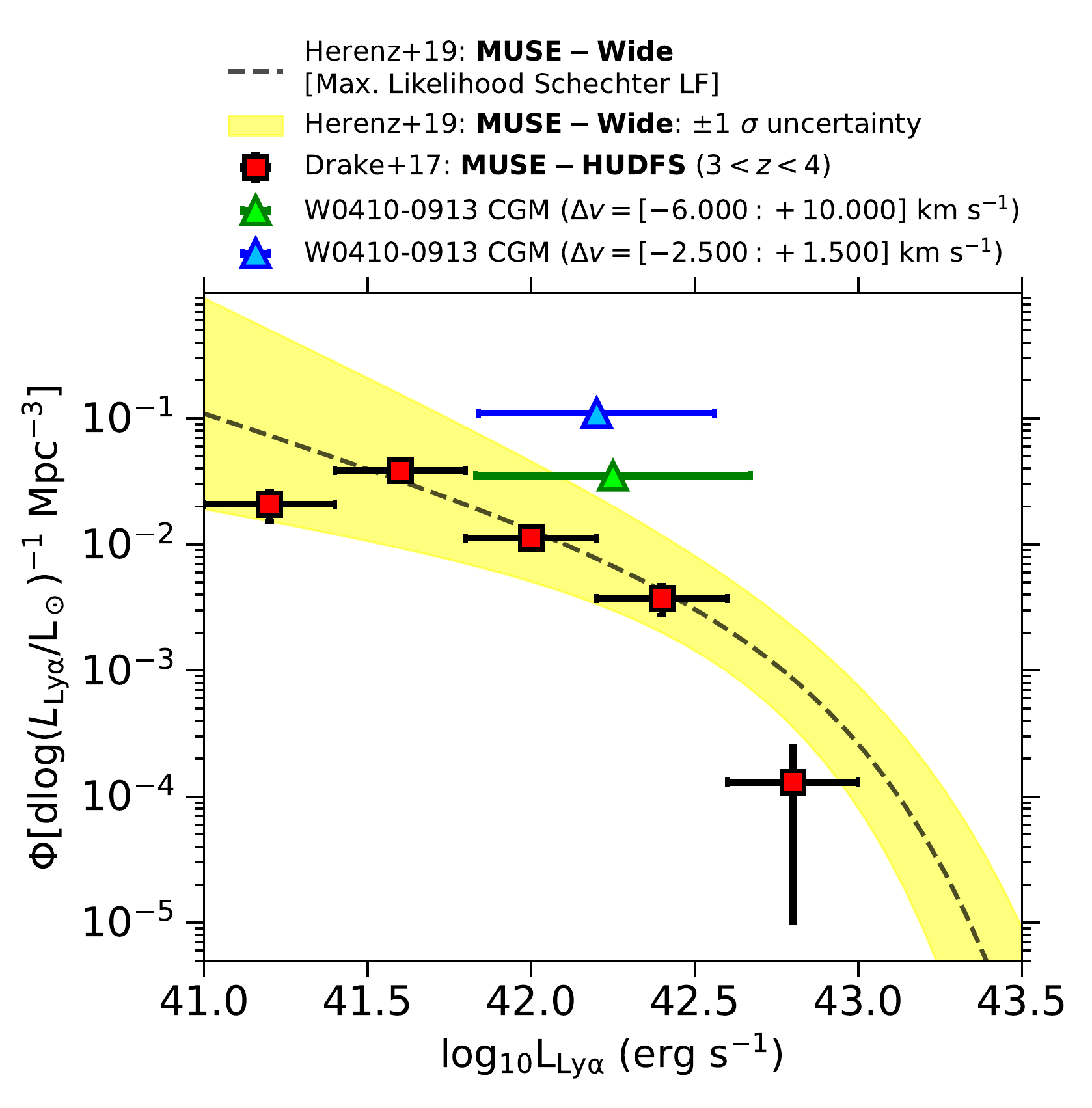}
	\caption{
		{\bf The Hot DOG W0410-0913 evolves in an very dense environment.}	
		We quantify the significance of the overdensity of companion galaxies discovered around W0410-0913, both in the whole volume defined by the offset velocity range $[-6.000; +10.000] ~ \rm{km ~ s^{-1}}$ (green triangle) and in the denser volume defined by the offset velocity range $[-2.500; +1.500] ~ \rm{km ~ s^{-1}}$ (blue triangle).
			We compare their Lyman-$\alpha$ luminosity-weighted number densities with the expected values for field galaxies at comparable redshift according to the Lyman-$\alpha$ luminosity functions measured with MUSE in wide \cite{Herenz2019} (gray line and yellow shaded region, representing the $\pm 1\sigma$ uncertainty on the fitting parameters) and deep \cite{Drake2017} surveys (red points and black lines, representing the $\pm 1\sigma$ errors).
		The number density of Lyman-$\alpha$-emitting galaxies, $\rm n_{LyE}$, associated with W0410-0913 is a factor of about 14 higher than the value ($\rm< n_{LyE} >$) expected for field galaxies with comparable Lyman-$\alpha$ luminosity. 
		The green and blue error bars represent the range of Lyman-$\alpha$ luminosities covered by companion galaxies in the whole and denser volume, respectively.
	}
	\label{fig:2}
\end{figure*}

To quantify the physical significance of the revealed overdensity we compare the Lyman-$\alpha$ luminosity-weighted number density of the companion galaxies associated with W0410-0913, $\rm n_{LyE}$, with the expected value for field galaxies according to the Lyman-$\alpha$ luminosity functions measured in wide and deep surveys with MUSE \cite{Drake2017,Herenz2019} (see subsection The physical significance of the overdensity in Methods for details on the computation and on the comparison sample). 
We find that the observed $\rm n_{LyE}$  is a factor of $\rho_{\rm CGM} \sim14$ higher than the expected value ($\rm< n_{LyE} >$; see Figure \ref{fig:2}).
This is higher than other reported observations of overdensities previously discovered through Lyman-$\alpha$ in the redshift range $z\sim 2-7$ \cite{Hayashino2004, Badescu2017, Bacon2021, Hu2021},  including proto-clusters around radio-galaxies \cite{Venemans2005,Venemans2007} and multiple-quasar objects \cite{Hennawi2015}, although a direct comparison with these systems is challenged by the differences in selection criteria, cosmological volumes, and sensitivity limits.
Also, we note that the measured overdensity is consistent with cosmological predictions in the proximity of massive, luminous objects like W0410-0913, at the redshift of interest (see subsection The physical significance of the overdensity in Methods).

Altogether, our findings suggest that the circumgalactic environment surrounding W0410-0913 (probed by MUSE on projected physical scales of approximately $400$ kpc) is one of the densest regions of the observed Universe. 
The number of gravitational encounters is expected to be enhanced in such a dense environment \cite{Hopkins2008, Fan2016, Farrah2017, Ginolfi2019}, suggesting that galaxy mergers could be an efficient channel for the mass build-up of the massive luminous galaxy.

However, our MUSE observations cannot firmly confirm this scenario, since we do not detect direct unequivocal signs of mergers.
A tentative indication of ongoing dynamical interactions is provided by the morphological and kinematic distribution of gas along the line-of-sight as traced by the Lyman-$\alpha$ emission in and around the companion galaxies, that we study by using commonly adopted flux-extraction techniques that enable one to recover flux of extended sources allowing for local kinematic variations \cite{Cantalupo2019}.
Specifically, we find that the three-dimensional (3D) distribution of circumgalactic gas illuminated by the Lyman-$\alpha$-emitters dynamically associated with the central luminous galaxy (i.e., with an offset velocity range of about $[-2.500; +1.500] ~ \rm{km ~ s^{-1}}$) is preferentially distributed over bridge-like shapes on scales up to $\gtrsim 100 ~{\rm kpc}$, some of which appear to be aligned along the common axis with the central massive object (see Figure \ref{fig:3}).
While these filamentary structures might be induced by the strong gravitational influence of W0410-0913, projection effects combined with the resonant nature of Lyman-$\alpha$ challenge this interpretation.

Further studies of the morphology and kinematics of the distribution of gas in the halo, tracing e.g., bright rest-frame FIR lines (like [CII]) with ALMA and rest-frame optical / near-IR lines with JWST, are needed to discern whether W0410-0913  is in the process of accreting its neighbors, similarly to what is done in highly interacting dense groups of massive galaxies in the local Universe through maps of atomic hydrogen filaments \cite{Yun1994}.

Furthermore, the merging scenario is at odds with high resolution spectroscopic imaging data of the internal gas structure in the ISM of the hyper-luminous central object.

\subsubsection*{ALMA observations of the internal ISM structure}

W0410-0913 has been extensively observed by ALMA targeting several rest-frame FIR lines in different bands.
While existing ALMA observations cannot provide information on the companion galaxies, they succeed in describing several properties on scales of the ISM, complementing the information retrieved by MUSE on the CGM environment.
Studying the emission from the rotational transition of carbon monoxide CO(6-5), observed in Band-4 with a high resolution of about 0.2$''$ (corresponding to $1.4~{\rm kpc}$ at this redshift), we find a smooth velocity gradient (see a map of the line-of-sight velocity in Figure \ref{fig:4}a) with velocities ranging between approximately $\rm  \pm ~500 ~ km ~ s^{-1}$. 
We perform a kinematic model of W0410-0913 using the 3D tilted ring model fitting code $^{\rm{3D}}$Barolo (see details on the procedure in Methods) and obtain models of the rotational velocity ($v_{\rm rot}$) and velocity dispersion ($\sigma$) radial profiles across the whole CO(6-5) emitting region (Figure \ref{fig:4}a).
The $v_{\rm rot}$ and $\sigma$ radial profiles are shown in Figure \ref{fig:4}b.
We find that, after a declining trend in the inner $ 1-2 ~ {\rm kpc}$, $v_{\rm rot}(r)$ follows a flat curve (within the uncertainties) reaching a value of about  $\rm  500 ~ km ~ s^{-1}$ at a radius of $\sim 6 ~ {\rm kpc}$.
The $\sigma(r)$ profile instead peaks at the center (about $\rm  200 ~ km ~ s^{-1}$ in the inner $\sim 1-2 ~ {\rm kpc}$) and slowly declines outward, exhibiting high values of $\rm \sigma \gtrsim 100 ~ km ~ s^{-1}$ up to about $5-6 ~ {\rm kpc}$.
In the lower panel of Figure \ref{fig:4}b we show, as a function of the radius, the ratio between rotational velocity and velocity dispersion, $v_{\rm rot}/\sigma$, which is typically used as an indicator of whether a system can be gravitationally supported by rotation or by turbulence \cite{Rizzo2020}.
We find that, at each radius, the rotational velocity is higher than the dispersion velocity, with an average (median) value of $v_{\rm rot}/\sigma = 3.5 \pm 1.8$ ($v_{\rm rot}/\sigma = 3.0 \pm 0.5$).
While W0410-0913 appears to be more turbulent than a few recently reported extreme dynamically cold disks at $z>3$ \cite{Rizzo2020, Neeleman2020, Lelli2021} (that show a surprisingly large $v_{\rm rot}/\sigma \sim 10$), the $v_{\rm rot}/\sigma$ values that we estimate are consistent with state-of-art galaxy formation models \cite{Pillepich2019, Kretschmer2021}, indicating that the molecular gas emission observed in our target arises from a massive rotationally supported fast rotating disk.

\subsection*{Discussion}

ALMA observations of W0410-0913, zooming into the ${\rm kpc}$-scale properties of its ISM, interpreted in the context of the CGM-scales studied with MUSE, challenge the scenario, suggested by models, that repeated galaxy encounters in the dense environment might have been efficient in building up the central galaxy. 
Indeed, multiple mergers in luminous systems are often shown to have a disruptive impact on the kinematics of the ISM \cite{Engel2010, Fan2016, Farrah2017}, which, in the case of W0410-0913, appears to be settled into a smooth rotation-dominated structure.

We speculate that a possible mechanism capable of solving this tension is that the mass growth of W0410-0913 could be sustained by minor mergers with relatively less massive companions, that do not significantly perturb the disk-like structure \cite{Kohandel2019, Kretschmer2021}.
In this case, the accretion of discrete small satellites might have played a role in originating the observed high velocity dispersion in W0410-0913 through gravitational instabilities.
Also, minor mergers are usually invoked by models as an efficient channel of bulge formation at the core of massive disk galaxies \cite{Kannan2015}. In the case of W0410-0913, the inner peak ($< 1-2 ~ {\rm kpc}$) of the CO(6-5) $v_{\rm rot}(r)$ and the high central concentration of flux in the intensity map could be interpreted as a tracer of a bulge-like structure, that minor mergers might have contributed to form.
However, we note the existence of an alternative (but not exclusive) scenario in which the energy injected by the prominent AGN activity into the surrounding gas might contribute to the observed turbulence throughout the disk, as suggested by similar observations of turbulent ISM revealed through spectroscopic imaging of molecular gas in another distant hyper-luminous obscured quasar \cite{DiazSantos2016, DiazSantos2018}.
\newline

Altogether, the multi-wavelength and multi-scale observations presented here, obtained with some of the most powerful telescopes available (including the first observations targeting a Hot DOG with MUSE) provide key information on the cosmic assembly of a distant hyper-luminous hot-dust obscured galaxy.
Data support that the mass build-up of W0410-0913 takes place in a dense circumgalactic environment, but we cannot identify robust signs of merger that could sustain the growth of the massive galaxy.
Instead, we find that gravitational encounters, if any, can have only a modest impact on shaping the morpho-kinematic properties of the ISM.
Our results serve as a benchmark to calibrate models of the rare and short-lived populations of Hot DOGs and progress in the understanding of their evolutionary transition towards luminous quasars and giant galaxies at cosmic noon.


\section*{Methods}

\subsection*{Cosmology }

Throughout the paper, we adopt a standard $\rm \Lambda$ cold dark matter ($\rm \Lambda CDM$) cosmology, using the following parameters compatible with Planck results \cite{Planck2018}: 
$\Omega_{\Lambda0} = 0.69$, $\Omega_{m0} = 0.31$, $\Omega_{b0} = 0.05$, $\rm H_0 = 67.7~ km ~s^{-1} ~Mpc^{-1}$.
Based on these parameters, at the redshift of W0410-0913 ($z=3.631$), the angular scale is $1'' = 7.39 ~{\rm kpc}$.

\subsection*{MUSE observations }

MUSE observations of W0410-0913 were carried out in February 2019 and January 2021, during the ESO Observing Periods P102 (Program 0102.A-0729) and P106 (Program 106.2184.001), respectively (see Data Availability Section).
The collection of data in P102 was organized in one observing block (OB) composed of four exposures of $\rm 680 ~s$ each, for a total exposure time on target of $\rm 45 ~m$ ($1 \rm ~h$ of telescope time including overheads). 
Seeing conditions ranged between ${\rm FWHM} = 0.8'' - 1.1''$, with airmass values of $1.1 - 1.3$.
In P106, the data acquisition was split into three exposures of $\rm 900 ~s$ each, with time on target ($\rm 45 ~m$) and telescope time ($1 \rm~ h$) similar to the P102 program. Observations were carried out under good seeing conditions, ${\rm FWHM} = 0.8''$, and airmass $\lesssim1.1$.
In both programs the exposures were slightly dithered by $\pm 0.3''$, and rotated with respect to each other by 90 degrees, to reduce systematic noise.

We have processed the individual exposures from both datasets with standard data reduction techniques using the \texttt{ESO-MUSE} Pipeline Version 2.8.4 \cite{Weilbacher2020}.
In detail, for each exposure, we used the  \texttt{MUSE\_BIAS}, \texttt{MUSE\_FLAT} and \texttt{MUSE\_WAVECAL} pipeline recipes \cite{Weilbacher2020} to perform bias subtraction, flat fielding, twilight and illumination correction, and wavelength calibration.
In order to ensure an accurate relative astronomy, we registered the individual exposures using the position of point sources in the field. This step is needed since the spatial shifts introduced by the derotator wobble \cite{Bacon2015} can cause offsets of a few arcsecs.

We performed the next stages of data reduction using the tools provided by the \texttt{CubExtractor} software package \cite{Cantalupo2019} (Version 1.8), which has been specifically developed to improve the data quality for the detection of diffuse, low-SB emission in MUSE datacubes \cite{Borisova2016, ArrigoniBattaia2019, Umehata2019, Fossati2021}.
In detail, we improved the flat fielding by using the tool \texttt{CubeFix} \cite{Cantalupo2019}, which performs an additional correction to homogenize the illumination across the field and as a function of wavelength.
We performed the sky subtraction using \texttt{CubeSharp} \cite{Cantalupo2019}, which can reach a better accuracy than the MUSE pipeline \cite{Borisova2016}.
We then combined the corrected and sky-subtracted individual exposures from both datasets using an average 3$\sigma$-clipping.
The final datacube samples the instrument FOV of $1'\times1'$ in pixels of size $0.2''$.
Each pixel contains a spectrum covering the wavelength domain $4750-9350$~\AA, and individual channels have a spectral resolution of $1.25$~\AA. 

Our combined MUSE observations, comprising a total telescope time of $2 ~ \rm{h}$, reach a ($2 \sigma$) SB limit of $5\times10^{-19} ~\rm{erg ~ s^{-1} cm^{-2} arcsec^{-2}}$ in $1 ~ \rm{arcsec^2}$ aperture and in a single spectral channel ($1.25 ~ \AA$), at wavelengths adjacent to $5632 ~\AA$, corresponding to the expected Lyman-$\alpha$ in W0410-0913 (systemic redshift of $z=3.631$; see next Sections).
For reference, the SB limit on a pseudo-narrow band (NB) image of $30 ~\AA$ (a common size for filters used in past NB surveys), obtained by collapsing 24 wavelength layers close to the location of Lyman-$\alpha$, is  $2.8\times10^{-18}~\rm{erg ~ s^{-1} cm^{-2} arcsec^{-2}}$ in $1 ~ \rm{arcsec^2}$ (2$\sigma$).

\subsection*{The rest-frame UV spectrum of W0410-0913}

In Figure \ref{fig:MUSEspectrum}a, we show the observed rest-frame UV spectrum of W0410-0913, covering the whole MUSE wavelength domain ($\sim 1000 - 2000 ~\AA$ rest-frame), extracted from a circular aperture with a radius of $1''$ (which maximizes the SNR) centered at the position of the luminous galaxy, based on its MUSE continuum image (obtained collapsing the full datacube). 
The spectrum shows a very faint rest-frame UV emission, 
with an average flux of $4.0 ~(\pm 3.2) \times10^{-19}~\rm{erg ~ s^{-1} cm^{-2} \AA^{-1}}$ computed at the rest-frame wavelength of $1300~ \AA$.
This corresponds to a monochromatic luminosity of ${\rm log}(\lambda L_\lambda / {\rm [erg~s^{-1}]}) = 43.8 ^{+0.3}_{-0.7}$, which is orders of magnitudes lower than the expected value ${\rm log}(\lambda L_\lambda / {\rm [erg~s^{-1}]}) = 47.3^{+1.1}_{-1.1}$, estimated through a standard bolometric correction relation for rest-frame UV luminosities \cite{Runnoe2012}, using $L_{\rm bol} = 1.68 \times 10^{14}  ~ {\rm L_\odot}$ for W0410-0913, as computed from multiwavelength photometry \cite{Tsai2015}.
This faint UV emission can be explained through dust attenuation caused by the large amount of obscuring ISM in the system \cite{Fan2016, Fan2018, DiazSantos2021}.

At the systemic redshift of W0410-0913, $z = 3.631$, traced by its molecular gas emission, 
we do not find any Lyman-$\alpha$ emission. 
The average flux, computed within a spectral window of $\Delta v = 1500 ~ {\rm km ~ s^{-1}}$ centered around the expected line center at $5632 ~\AA$, is consistent with zero within the uncertainties, $\sim 3 ~(\pm 30)\times10^{-20}~\rm{erg ~ s^{-1} cm^{-2} \AA^{-1}}$ (see Figure \ref{fig:MUSEspectrum}b), and by producing pseudo-NB images at the expected position of Lyman-$\alpha$ (collapsing the continuum-subtracted datacube along the wavelength axis using different spectral widths, from $5$ to $30~\AA$) we do not find any sign of emission.
However, a narrow emission line ($\rm FWHM \sim  400 ~(\pm 80) ~ km ~ s^{-1}$), emitted at the same location of W0410-0913, is detected in the spectrum with a blueshift of about $\rm  - 1500 ~(\pm 150) ~ km ~ s^{-1}$ with respect to $z = 3.631$ (see  Figure \ref{fig:MUSEspectrum}). 
We interpret this line as a Lyman-$\alpha$ emission from W0410-0913, emitted at different velocities than the systemic redshift traced by the molecular gas.
This interpretation is supported by the blueshift of about $\rm  - 1400 ~ km ~ s^{-1}$ (with respect to the systemic redshift adopted in this work) of the narrow component of the non-resonant line \oiii \l  5007 from a rest-frame optical spectrum of W0410-0913 observed with KECK/NIRES \cite{Finnerty2020}.
These findings suggest that the rest-frame UV and optical narrow emission lines in W0410-0913 are emitted at different velocities than the molecular gas.
An alternative, but less likely (because of the reasons discussed above) interpretation is that the narrow rest-frame UV emission line that we find in the MUSE spectrum traces Lyman-$\alpha$ emission produced by a nearby companion galaxy that is part of the overdense environment (see subsection The circumgalactic environment of W0410-0913 observed with MUSE), and is located along the line of sight of W0410-0913.

We compute the spatial extension of such emission using the tool \texttt{CubEx} (from the \texttt{CubExtractor} package \cite{Cantalupo2019}).
\texttt{CubEx} is extensively used to perform automatic detection and extraction of extended sources in MUSE datacubes \cite{Borisova2016, ArrigoniBattaia2019}, and it is based on a 3D extension of the connected-labeling-component algorithm with union finding, usually used in classical binary image analysis \cite{Borisova2016}.
We feed \texttt{CubEx} with a sub-cube that we extract to have a wavelength range of $\Delta \lambda \sim 100~\AA$ centered around the peak of Lyman-$\alpha$ emission ($4388~\AA$), and from which we remove continuum sources by adopting a median-filtering approach.
We also apply a spatial Gaussian filtering of 2 pixels ($0.4''$) and a spectral smoothing with a Gaussian filter size of 1 layer ($1.25~\AA$) to bring out extended but narrow features.
Running \texttt{CubEx} with a SNR threshold for detection of ${\rm SNR} > 3$ in individual datacube elements (voxels) we detect a nebula consisting of approximately $1600$ connected voxels and a maximum spectral width, as defined in the 3D-segmentation-mask, of $\Delta \lambda \sim 15$~\AA.
From the optimally extracted image of the nebula, derived by applying the 3D-segmentation-mask to the datacube (see subsection A study of the 3D distribution of gas around W0410-0913, for details on the advantage of this method), we measure a spatial extension, defined as the maximum projected linear size, of about $30 ~\rm kpc$ (see Figure \ref{fig:MUSEnebula}).
This indicates that, under the reasonable assumption that such emission could be considered a Lyman-$\alpha$ nebula associated with the Hot DOG, it would be smaller than several nebulae observed with MUSE (usually with lower exposure times) around UV-bright quasars with lower $L_{\rm bol}$ \cite{Borisova2016, ArrigoniBattaia2019}, pointing toward a significant effect of a dusty ISM on the detectability of Lyman-$\alpha$ nebulae.

In the MUSE spectrum of W0410-0913 we also clearly detect the \nv \l 1240 line emission and the \civ \ll 1548,1550 doublet, both showing very broad line-profiles with $\rm FWHM \sim  2800 ~(\pm 200) ~ km ~ s^{-1}$, computed through a non-linear least-squares fit of a Gaussian profile. 
Similarly to other luminous quasars and other Hot DOGs \cite{DiazSantos2018}$^,$ \cite{Marziani2017}, by computing the center of their Gaussian profiles, we measure large blueshifts for both \nv  and  \civ of about $\rm  2000 ~(\pm 150) ~ km ~ s^{-1}$ (see Figure \ref{fig:MUSEspectrum}), corresponding to a redshift of $z = 3.600 \pm 0.005$.
This value is consistent with the first published estimate of the W0410-0913 optical redshift, based on shallow observations of \civ and \Heii \ll 1640 emission obtained with Keck/LRIS \cite{Wu2012}.
We note that the MUSE spectrum does not show the low SNR \Heii emission present in the Keck/LRIS, as the line falls in a noisy region of the spectrum.
Furthermore, similarly to the \nv and \civ broad lines in the MUSE spectrum, the optical \oiii observed with KECK/NIRES (see above) has a broad component that is also very blueshifted, showing an offset velocity of $\rm  3000 ~(\pm 1000) ~ km ~ s^{-1}$ with respect to the narrow component \cite{Finnerty2020}.
Large blueshifts in the rest-frame UV and optical broad emission lines, like the ones observed in W0410-0913, are often interpreted as an indication of fast nuclear outflows \cite{Vietri2018}.

\subsection*{The  circumgalactic environment of W0410-0913 observed with MUSE}

To study the environment of W0410-0913 over an area of about ${\rm 450 ~kpc \times 450 ~kpc}$ (corresponding to the MUSE $1' \times 1'$ FOV), we search for Lyman-$\alpha$-emitting companion galaxies in a continuum-subtracted sub-cube (produced through median-filtering) of about $380 ~\AA$ centered around the expected Lyman-$\alpha$ emission from W0410-0913, corresponding to offset velocities of $[-10.000; +10.000] ~ \rm{km ~ s^{-1}}$ with respect to its systemic redshift.

Our method can be summarized as follows. We inspect the whole wavelength domain by producing mini-cubes with spectral widths ranging from a minimum allowed number of 2 channels ($2.5~\AA$; to avoid single-channel spikes of noise) to a maximum of 20 channels (about $25 ~\AA$; to allow for $\sim 3\times$ typical FWHMs of Ly$\alpha$-emitters at this redshift).  
We collapse each mini-cube along the wavelength direction to obtain pseudo-NB images that we scan through their whole spatial domain by measuring the aperture photometry of $1''-$sized regions centered around each pixel.  We consider as candidate detections the regions with a significance of their measured flux of ${\rm SNR}>3$ (where the noise is extracted from the local background), we extract a spectrum, and we save their position (RA and DEC of the central pixel), central wavelength and spectral width. For multiple candidate detections arising from the same physical source, namely two or more outputs found at about the same location (within $0.5''$, corresponding to about half of the size of the PSF) and wavelength (within $4 ~\AA$), we keep the candidate that has higher SNR in the spectrum.
We then perform a visual inspection of the pseudo-NB images and spectra of the candidate detections to make sure that there is no contamination by sky residuals. 
Also, for each candidate detection, we visually inspect the original non-continuum subtracted full datacube (covering the whole MUSE spectral domain) to check whether the candidate detections show the clear presence of multiple (at least two) emission lines at wavelengths not included in the sub-cube used for studying the environment of W0410-0913. 
In these cases, the sources would be line emitters at lower redshifts (emitting e.g., [OII] \l 3727,3729,  H$\beta$, [OIII] \l 4959,5007 and H$\alpha$) and would not be relevant for our analysis.
Through this analysis we exclude from the list two sources emitting [OII] at $z = 0.495$ (H$\beta$ and the [OIII] doublet are detected at redder wavelengths) and H$\beta$ at $z= 0.167$ (with [OII], [OIII] and H$\alpha$ detected in the full spectrum).
We assume that the remaining sources, showing a single emission line in the sub-cube centered around the redshift of W0410-0913 are Lyman-$\alpha$-emitting galaxies in the surroundings of W0410-0913.

We find 24 Lyman-$\alpha$ emitters within a box with a diameter of about $400 ~{\rm kpc}$ centered around W0410-0913.
This result is confirmed by an independent analysis that we perform using \texttt{CubEx}. Although \texttt{CubEx} is usually exploited to extract diffuse extended emission, it can be used to mimic a line finder by decreasing the threshold of the minimum allowed number of connected voxels and by increasing the threshold of SNR for the detection of individual voxels. 
In detail, by adopting a $SNR  > 4$ threshold over a lightly smoothed cube ($0.2''$ pixels in the spatial domain, and no filtering on the wavelength direction), we find the same number of sources of the method described above, with only tiny differences ($\lesssim 0.3''$) in the spatial location of the peak of a few objects, likely due to the different pre-processing of the cube.

In Supplementary Figures 1 and 2, we show the spectra of these sources (extracted from a region with a radius of $1''$) centered around the central wavelength of their Lyman-$\alpha$, that we compute through a non-linear least-squares Gaussian fit. 
A selection of these spectra is shown in Figure \ref{fig:selection_companions}a.
From the modeled Lyman-$\alpha$ peaks we measure the offset velocities with respect to the redshift of W0410-0913, and find values in the range $[-6.000; +10.000] ~ \rm{km ~ s^{-1}}$, spatially distributed in the MUSE FOV as shown in Figure \ref{fig:1}b. 
We note that radiative transfer effects are known to shift the peak of Lyman-$\alpha$ emission in the spectra of star-forming galaxies even by a few hundreds of $\rm{km ~ s^{-1}}$ compared to the systemic velocity \cite{Verhamme2018}, though this effect is unlikely to have a dominant impact on the suggested interpretation of the observations, since we measure offset velocities up to thousands of $\rm{km ~ s^{-1}}$.

We also produce pseudo-NB images of each object by collapsing the datacube across the wavelength direction within spectral ranges defined by ${\rm [\lambda_{cen} - FWHM: \lambda_{cen} + FWHM]}$, where the central wavelength ${\rm \lambda_{cen}}$ and ${\rm FWHM}$ are defined by the spectra shown in Supplementary Figures 1 and 2.
A composition of the derived pseudo-NB images, showing the projected spatial distribution of the Lyman-$\alpha$-emitting galaxies around W0410-0913, is shown in Figure \ref{fig:1}a.

As shown in Figure \ref{fig:1}c, the distribution of offset velocities is centered around W0410-0913, confirming that the massive luminous galaxy is located at the dynamic center of its dense environment and dominates its gravitational potential. 
Most of the companions (18 out of 24) show offset velocities in a smaller range of about $[-2.500; +1.500] ~ \rm{km ~ s^{-1}}$. 
With the caveat of partial information provided by the measurement of projected line-of-sight velocities, it is likely that these galaxies are more closely interacting with the central massive object.

We dub the Lyman-$\alpha$-emitters using alphabetically ordered letters from $\rm \#A$ to $\rm \#X$, according to their velocity offsets, from the largest negative to the largest positive.
We compute the Lyman-$\alpha$ fluxes of the companion galaxies by integrating the Gaussian model of their spectra, and convert them to Lyman-$\alpha$ luminosities ($L_{\rm Ly\alpha}$), obtaining values in the range ${\rm log}(L_{\rm Ly\alpha} / {\rm [erg~s^{-1}]}) \sim 41.8 - 42.65$.
A summary of the measured properties of the Lyman-$\alpha$-emitters is reported in Table 1.

\subsection*{A statistical assessment of the overdensity}

To provide a statistical assessment of our finding, we perform the same analysis discussed in the previous subsection on other four $380 ~\AA$-sized sub-cubes adjacent to the sub-cube used to study the environment of W0410-0913, in the large wavelength range  $4750 - 6500 ~\AA$ (corresponding to offset velocities of $[-50.000; +50.000] ~ \rm{km ~ s^{-1}}$). 
This wavelength range is optimal because the sensitivity of the instrument varies only by a factor less than $25\%$, and it is not strongly affected by sky lines emission (see a description of MUSE performances at \url{https://www.eso.org/sci/facilities/paranal/instruments/muse/inst.html}).
After checking for contamination from sky-residuals and multiple line emitters at intermediate redshifts (the number of the latter is found to be between 0 and 2 per sub-cube), we find that the number of detected candidate Lyman-$\alpha$ emitters is either 1 and 2, 
which confirms the overdense nature of the environment in the surrounding of the Hot DOG.

As an additional sanity check, using the same method, we scan through the negative version (obtaining by changing the sign to each pixel) of the sub-cubes defined above.
In this case the visual inspection following the line finder algorithm is limited to a check of the contamination by sky-residuals.
We find that the number of detections is 0 in all the sub-cubes. 
This excludes the possibility that a fraction of the positive sources could be resulting from spikes of noise and thus be spurious.

\subsection*{The physical significance of the overdensity}

To quantify the significance of the discovered overdensity we compute the Lyman-$\alpha$ luminosity-weighted number density ($\Phi$) of Lyman-$\alpha$-emitting galaxies surrounding W0410-0913 and compare it with the value predicted by the most updated Lyman-$\alpha$ luminosity functions measured for field galaxies in wide and deep surveys with MUSE \cite{Drake2017,Herenz2019}.

Given the non-uniform and highly peaked distribution of offset velocities (see Figure \ref{fig:1}c), we derive two values of $\Phi$, distinguishing between the whole volume containing 24 Lyman-$\alpha$ emitters (including W0410-0913) in the velocity range $[-6.000; +10.000] ~ \rm{km ~ s^{-1}}$ ($\Phi_{\rm whole}$) and a smaller (but denser) volume containing 18 Lyman-$\alpha$ emitters in the velocity range $[-2.500; +1.500] ~ \rm{km ~ s^{-1}}$ ($\Phi_{\rm dense}$).

The $1' \times 1'$ projected area (i.e., the MUSE FOV) over which we find the companion galaxies corresponds to a comoving area of $4.2 ~{\rm Mpc^2}$.
In the whole volume case, the {\rm redshift-side} has a comoving size of about $190~ {\rm Mpc}$, which yields to a cosmological comoving volume of $800 ~{\rm Mpc^3}$ and a number density of Lyman-$\alpha$ emitters of $n_{\rm LyE} \sim 3 \times 10^{-2}~{\rm Mpc^{-3}}$.
In the dense volume case, the {\rm redshift-side} has a smaller comoving size of $50~ {\rm Mpc}$, yielding to a cosmological comoving volume of about $210 ~{\rm Mpc^3}$ and a number density of Lyman-$\alpha$ emitters of $n_{\rm LyE} \sim 8.5 \times 10^{-2}~{\rm Mpc^{-3}}$.
To convert $n_{\rm LyE}$ into $\Phi$ we divide by the range of Lyman-$\alpha$ luminosities of galaxies in both volumes,
namely $\Delta {\rm log}(L{_{\rm Ly\alpha}}/{\rm L_\odot}) = 0.86$ for the whole volume and $\Delta {\rm log}(L{_{\rm Ly\alpha}}/{\rm L_\odot}) = 0.75$ for the dense volume.
We obtain Lyman-$\alpha$ luminosity-weighted number densities of $\Phi_{\rm whole} = 3.5 \times 10^{-2} ~ {\rm dlog}(L{_{\rm Ly\alpha}}/{\rm L_\odot})^{-1} ~ {\rm Mpc^{-3}}$ and $\Phi_{\rm dense} = 1.1 \times 10^{-1} ~ {\rm dlog}(L{_{\rm Ly\alpha}}/{\rm L_\odot})^{-1} ~ {\rm Mpc^{-3}}$.
These values are higher than the expected values for field galaxies (see Figure \ref{fig:2}).

In the following we focus on the dense volume, which is more likely to contain companion galaxies gravitationally associated with W0410-0913.
In detail, at the median Lyman-$\alpha$ luminosity of galaxies in the dense volume, ${\rm log}(L{_{\rm Ly\alpha}}/{\rm L_\odot}) = 42.2$, the MUSE-wide survey \cite{Herenz2019}, which observed Lyman-$\alpha$-emitters over a large area of $>20~ {\rm arcmin}^2$, yields an expected $\Phi_{\rm field} = 10^{- (2.1^{+0.37}_{-0.47})} ~ {\rm dlog}(L{_{\rm Ly\alpha}}/{\rm L_\odot})^{-1} ~ {\rm Mpc^{-3}}$, where the uncertainty represents the errors on the best-fitting Schechter function parameters \cite{Herenz2019}.
Comparing this value with $\Phi_{\rm dense}$, we find that the circumgalactic environment around W0410-0913 has a denser concentration of Lyman-$\alpha$-emitting galaxies than the average volume, with an estimated overdensity factor of $\rho_{\rm CGM}= \dfrac{\Phi_{\rm dense} }{\Phi_{\rm field} } = 14^{+16}_{-8}$.
We note that this calculation is conservative since it is based on the assumption that offset velocities with respect to the central object are tracing relative distances, while it is more likely that the observed dispersion of offset velocities (at least for objects in the dense volume) is affected by gravitational attraction (see Article and previous subsections). 
Thus the effective values of $\Phi$ in our field are likely to be higher, e.g., up to a factor of about $20$ times higher in the dense volume case under the extreme assumption that the 18 Lyman-$\alpha$ emitters are contained in a cube with a side of $60''$ centered around W0410-0913.

Despite its high value, we find that the measured overdensity of Lyman-$\alpha$ emitters in the dense volume, $\rho_{\rm CGM} = 14^{+16}_{-8}$, is consistent with cosmological predictions in the proximity of extreme, luminous objects like W0410-0913.

To check this, we first estimate how rare the fluctuation of matter density associated with $\rho_{\rm CGM}$ would be.
The first step is to consider that in general the spatial clustering of observable galaxies does not precisely mirror the clustering of the bulk of matter in the Universe.
This caveat is often addressed through the definition of a galaxy clustering bias, which describes the relationship between the spatial distribution of galaxies and the underlying dark matter density field \cite{Coil2013}.
Thus, by using a standard assumption of linear bias \cite{Kovac2007}, we convert the observed overdensity of Lyman-$\alpha$ emitters $\rho_{\rm CGM}$ into an estimate of the underlying matter overdensity, $\rho_{\rm M}$, by dividing it by the clustering bias parameter of Lyman-$\alpha$ emitters, $b$.
By using an updated value of $b=2.8\pm0.4$, recently computed through a 3D clustering analysis of hundreds of Lyman-$\alpha$ emitters at $z\lesssim4$ in the large MUSE-Wide survey \cite{HerreroAlonso2021}, and consistent with previous similar studies \cite{Kovac2007, Ouchi2018}, we obtain an estimate of the matter overdensity in the surroundings of W0410-0913 of $\rho_{\rm M} = 5^{+7}_{-3}$.
To estimate the rarity of such a fluctuation, we start from the computation of the linear rms density fluctuation\cite{Kravtsov2012}, $\sigma_{\rm LN}$, predicted by the standard $\rm \Lambda CDM$ cosmology. For the dense volume over which we compute $\rho_{\rm M}$, i.e., about $210 ~{\rm Mpc^3}$, we find a value of $\sigma_{\rm LN} \sim 1.6$.
The next step is to compute the probability associated with observing an overdensity $\rho_{\rm M}$ given a certain density fluctuation.
On the physical scales of our system, the density field is non-linear, and a possible approximation for the true density distribution $P(\rho)$ is a log-normal model \cite{Klypin2018}, for which the $\sigma_{\rm LN} \sim 1.6$ can be interpreted as the rms in $\rm ln(\rho)$, with a mean in this quantity of ~$ ~ - \dfrac{\sigma_{\rm LN}^2}{2}$. 
Under this assumption, a matter overdensity of  $\rho_{\rm M} = 5^{+7}_{-3}$ corresponds to a probability of $P(\rho_{\rm M}) = 10^{-1.4\pm 0.6}$. 
Given the considered volume of ~$210 ~{\rm Mpc^3}$ and the estimated $P(\rho_{\rm M})$, we find that the comoving density of objects like the central galaxy would need to be about  $n_{\rm halo} \sim \dfrac{P(\rho_{\rm M})}{210 ~{\rm Mpc^3} } \sim 10^{-3.7\pm0.6} ~{\rm Mpc^{-3}}$.
This values is consistent, within the uncertainties, to the predicted comoving density of halos hosting central massive galaxies like W0410-0913, with maximum velocities of $v_{\rm max} = 500 \pm 70 ~ {\rm km ~s^{-1}}$ (see the radial profile of the rotational velocity in Figure \ref{fig:4}b) at $z=3.631$, which we find to be $n_{\rm halo} \sim 10^{-3.8 \pm0.5} ~{\rm Mpc^{-3}}$, as we compute integrating the halo mass function provided by the large N-body simulation MDPL2 ~\cite{Klypin2016}, included in the open-access CosmoSim database (\url{https://www.cosmosim.org/cms/simulations/mdpl2/}).

\subsection*{Measure of the SFR from HST}

We estimate the SFR of the Lyman-$\alpha$-emitting companion galaxies around W0410-0913 using archival HST data, that we retrieve from the public Barbara A. Mikulski Archive for Space Telescopes (MAST) portal (\url{https://archive.stsci.edu/}).  
A $2' \times 2'$ field around our object was observed for a total exposure time of $2400~{\rm s}$ in October 2012 during the HST Cycle 20 (Program ID: 12930) exploiting the filter F160W on the Wide Field Camera 3 (WFC3), centered at about $15436 ~ \AA$, which corresponds to $\sim 0.3 ~{\rm \mu m}$ at the redshift of interest.

We extract the fluxes in the  HST WFC3/F160W image from circular apertures with $1''$-sized diameters centered around the peak emission, that we search within a radius of $0.5''$ from the Lyman-$\alpha$ peak (obtained through a 2D Gaussian fit of the pseudo-NB images; see Figure \ref{fig:1}a), to allow for slight spatial offsets between Lyman-$\alpha$ and the rest-frame UV/optical continuum \cite{Hoag2019}. 
We then derive the luminosity density $L_{\nu}$ and convert it in SFR by using a standard rest-frame UV/optical $L_{\nu} - {\rm SFR}$ calibration and assuming a Chabrier initial mass function \cite{Kennicutt1998}.
Although the available HST data are shallow (with a $3\sigma$ flux limit corresponding to a ${SFR}_{3\sigma} \sim 9 ~{\rm M_{\odot}~yr^{-1}}$), we detect 10 galaxies at ${\rm SNR} > 4$, with star formation rates ranging between ${\rm SFR} \sim 12 - 100 ~{\rm M_{\odot}~yr^{-1}}$.

In Supplementary Figure 3, we show the HST cutouts for the companion galaxies detected with the WFC3/F160W filter, along with the Lyman-$\alpha$ contours from MUSE.
A selection of HST maps is shown in Figure \ref{fig:selection_companions}b.
The SFRs are reported in Table 1.

To obtain a better constraint of the average star formation rate in the 14 companion galaxies not detected by HST we perform a median stacking of their $5''\times 5''$ cutouts centered at the Lyman-$\alpha$ peaks.
Prior to stacking, we run a sigma-clipping algorithm (with a $3\sigma$ threshold on individual pixels) over each cutout to mask possible continuum sources outside the central circular area of $1''$ diameter. 
This is necessary to guarantee an uncontaminated estimate of the noise level in the final image, that would be otherwise affected by bright field objects.
The noise level in the stacked image decreases down to a corresponding $3\sigma$ SFR limit of ${SFR}_{3\sigma} \sim 2.5  ~{\rm M_{\odot}~yr^{-1}}$, enabling us to measure with a ${\rm SNR} \sim 3.6$ an average ${\rm SFR} \sim 3 ~{\rm M_{\odot}~yr^{-1}}$, suggesting that a non-negligible star formation is taking place in the companion galaxies not detected by HST too.

\subsection*{A study of the 3D distribution of gas around W0410-0913} 

Extended diffuse Lyman-$\alpha$ emission, mainly produced through recombination radiation following photoionization (often referred to as fluorescence) powered by UV-bright sources\cite{Cantalupo2014}, is a commonly used tracer to observe the cool gas (with temperatures of approximately $T = 10^4~{\rm K}$) in the CGM around galaxies.
At $z \gtrsim 2 - 3$, where the Lyman-$\alpha$  is redshifted into the optical domain, great advances have been recently made with wide-field integral field spectrographs at large telescopes (like MUSE, or the Keck Wide-Field Imager), which have identified ubiquitous Lyman-$\alpha$ nebulae around quasars, extending up to hundreds of kpc \cite{Borisova2016, ArrigoniBattaia2019}, and Lyman-$\alpha$ halos around star-forming galaxies, extending up to tens of kpc, between factors of $5 - 20$ larger than the rest-frame UV sizes \cite{Leclercq2020}.

We search for extended Lyman-$\alpha$ emission around the companion galaxies of W0410-0913 to study the morphological and kinematic distribution of gas in the circumgalactic environment of W0410-0913.
We run \texttt{CubEx} over the sub-cube in which the 24 Lyman-$\alpha$-emitters are found, using the same pre-processing and setup of parameters adopted for analyzing the Lyman-$\alpha$ nebula (see previous subsections), namely a ${\rm SNR} > 3$ threshold for voxel-detection in a smoothed cube with a Gaussian spatial filter of $0.4''$ and a Gaussian spectral filter of $1.25~\AA$.
This setup is different from that exploited in the line-emitters searches, and it is tailored at extracting faint and narrow features.
We produce optimally extracted Lyman-$\alpha$ flux maps of the sources by selecting the voxels detected in their 3D-segmentation-masks and integrating their fluxes along the wavelength axis. 
This technique is operationally equivalent to combining different pseudo-NB images in which the size of the spectral filter is optimized for each pixel. 
The advantage of this approach, enabled by integral field spectrographs, is that one can maximize the SNR of the whole emitting region by revealing both kinematically narrow and broad features of a source at the same time. In contrast, imaging through classical NB filters with a fixed width would produce either a noise-dilution or an underestimation of the signal emitted.

\begin{figure*}[ht]
	\centering
	\includegraphics[width=\linewidth]{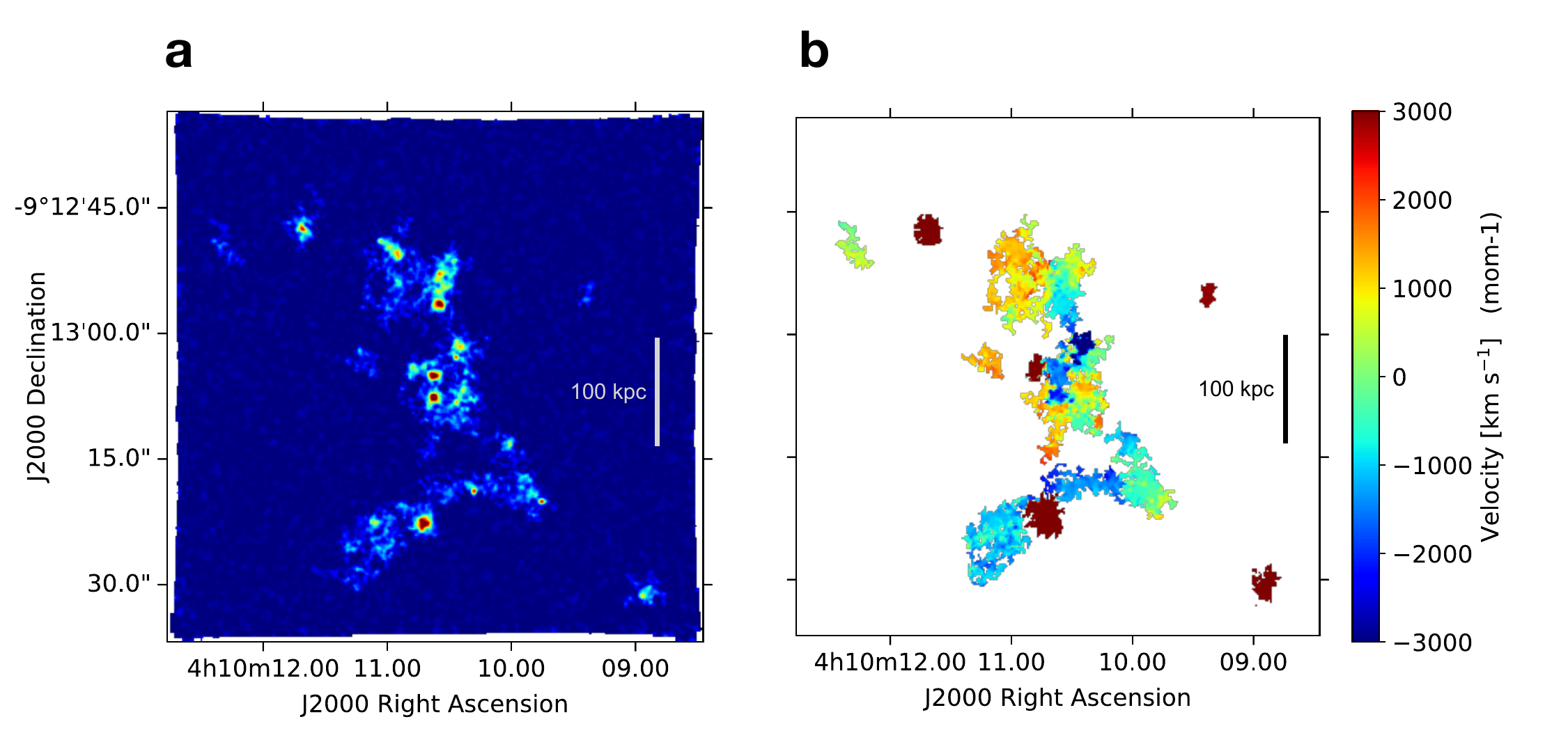}
	\caption{
		{\bf Morphologically disturbed and filamentary distribution of gas in the surroundings of W0410-0913.}	
		In panel {\bf a} and panel {\bf b} we show a composition of the optimally extracted Lyman-$\alpha$ flux maps and kinematic maps of galaxies in the overdensity, respectively, obtained by computing the moment-0 and moment-1 of the voxels detected in their 3D-segmentation-masks (see subsection A study of the 3D distribution of gas around W0410-0913 in Methods).
		The three-dimensional distribution of gas around galaxies in the denser dynamic volume (with offset velocities in the range $[-2.500; +1.500] ~ \rm{km ~ s^{-1}}$) 
		is suggestive of connecting filamentary structures on scales up to $\gtrsim 100 ~{\rm kpc}$ (see white and black scales).
	}
	\label{fig:3}
\end{figure*}

\begin{figure*}[ht]
	\centering
	\includegraphics[width=\linewidth]{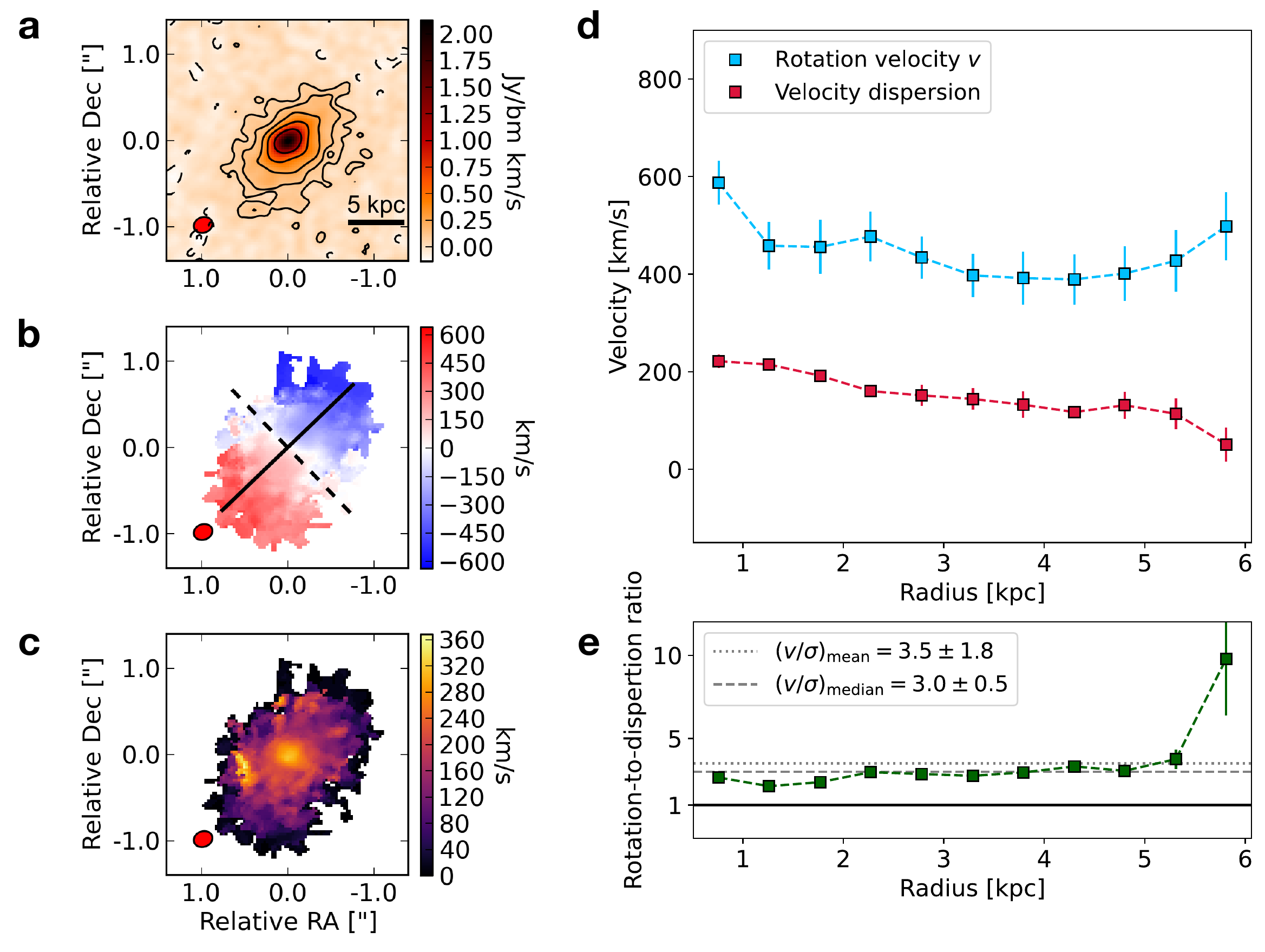}
	\caption{
		{\bf ALMA CO(6-5) observations of W0410-0913 show that its molecular ISM is structured in a rotationally-supported fast-rotating disk.}	
		In panel {\bf a} we show the moment-0 (velocity-integrated intensity), in panel {\bf b} the moment-1 (velocity field),
		and in panel {\bf c} the moment 2 (velocity dispersion field). 
		The black solid line in the lower-left corner of the moment-0 panel shows a 5 kpc physical scale.
		The solid line in the moment-1 represents the kinematic major axis, while the dashed line represents the minor axis.
		The black contours in the moment-0 represent the $\rm \pm [2, 4, 8, 16, 32] ~\sigma$ levels emission, where $\sigma = 3.3 \times 10^{-2} ~ {\rm Jy~beam^{-1}~km~s^{-1}}$ is the root mean square (rms) of the velocity-integrated CO(6-5) flux map.
		The ALMA beam, colored in red, is shown in the lower-left corners.
		We also show, as a function of radius, the profiles of rotational velocity (blue squares and lines) and velocity dispersion (red squares and lines) in panel {\bf d} and the profile of their ratio (green squares and lines) in panel {\bf e}. 
		The error bars represent the standard deviation derived by the model.
		The dotted (dashed) black line in the lower panel indicates the average (median) value of $v_{\rm rot} / \sigma$.
	}
	\label{fig:4}
\end{figure*}

\begin{figure*}[ht]
	\centering
	\includegraphics[width=\linewidth]{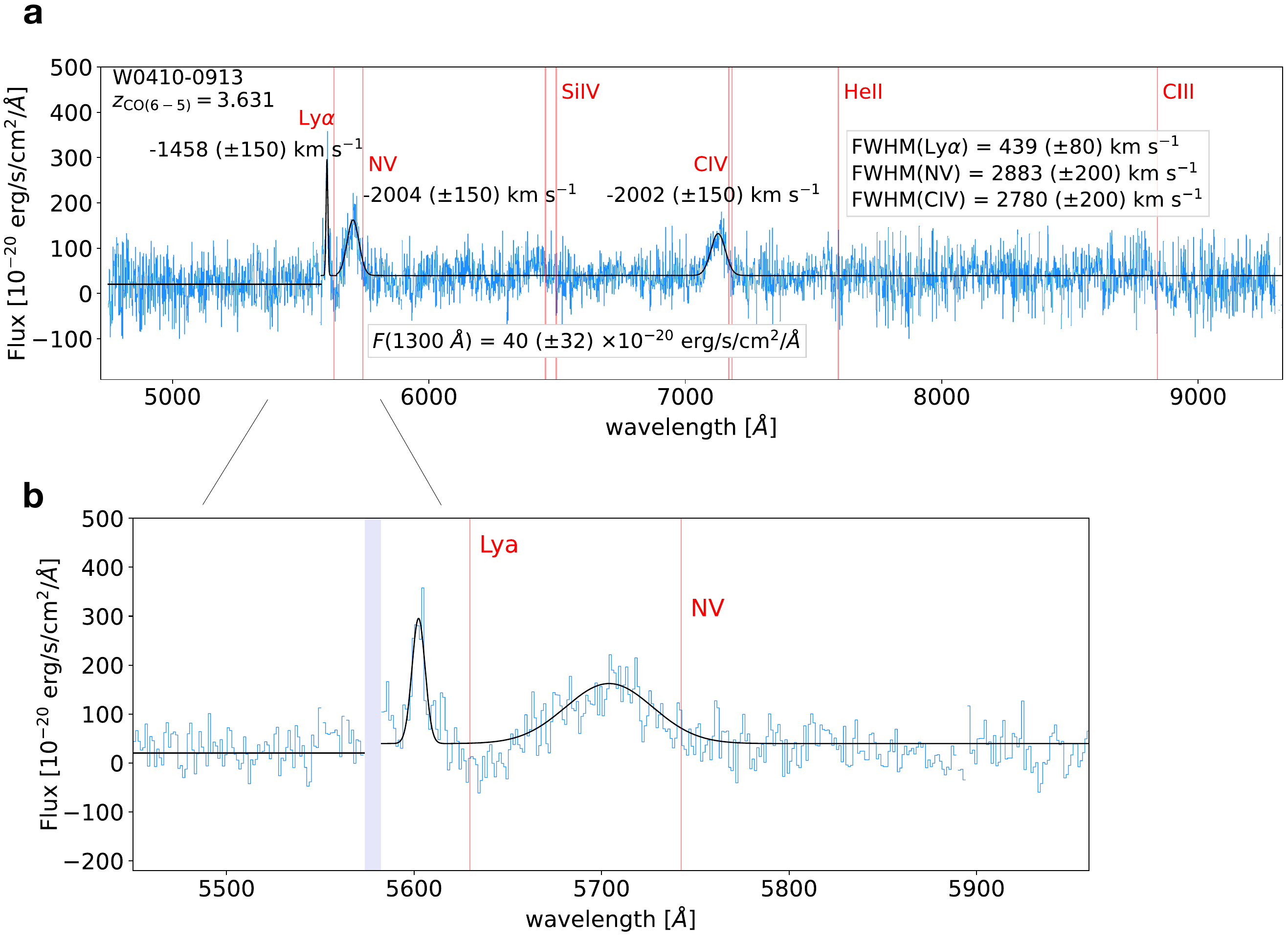}
	\caption{
		{\bf MUSE Spectrum of W0410-0913.}
		{\bf a} We show a spectrum of W0410-0913, covering the full MUSE wavelength domain, extracted from a circular aperture with a $1''$-radius.
		Vertical red lines are located at the expected position of several rest-frame UV lines, according to its redshift (as traced by the rest-frame FIR lines).
		A sigma-clipping procedure was performed to remove the residuals of sky-lines and make the visualization easier.
		\nv \l 1240 and \civ \ll 1548,1550 show very broad line-profiles ($\rm FWHM \sim  2800 ~ km ~ s^{-1}$) and large blueshifts ($\rm  -2000  ~ km ~ s^{-1}$) with respect to W0410-0913, indicative of fast nuclear outflows. 
			A narrow Lyman-{$\alpha$} line ($\rm FWHM \sim  400 ~ km ~ s^{-1}$) is detected with an offset of $\rm  \sim -1460  ~ km ~ s^{-1}$ from the systemic redshift.
		{\bf b} We show a zoom of the spectrum at wavelengths closer to the expected Lyman-{$\alpha$} line.
	}
	\label{fig:MUSEspectrum}
\end{figure*}

\begin{figure*}[ht]
	\centering
	\includegraphics[width=1\linewidth]{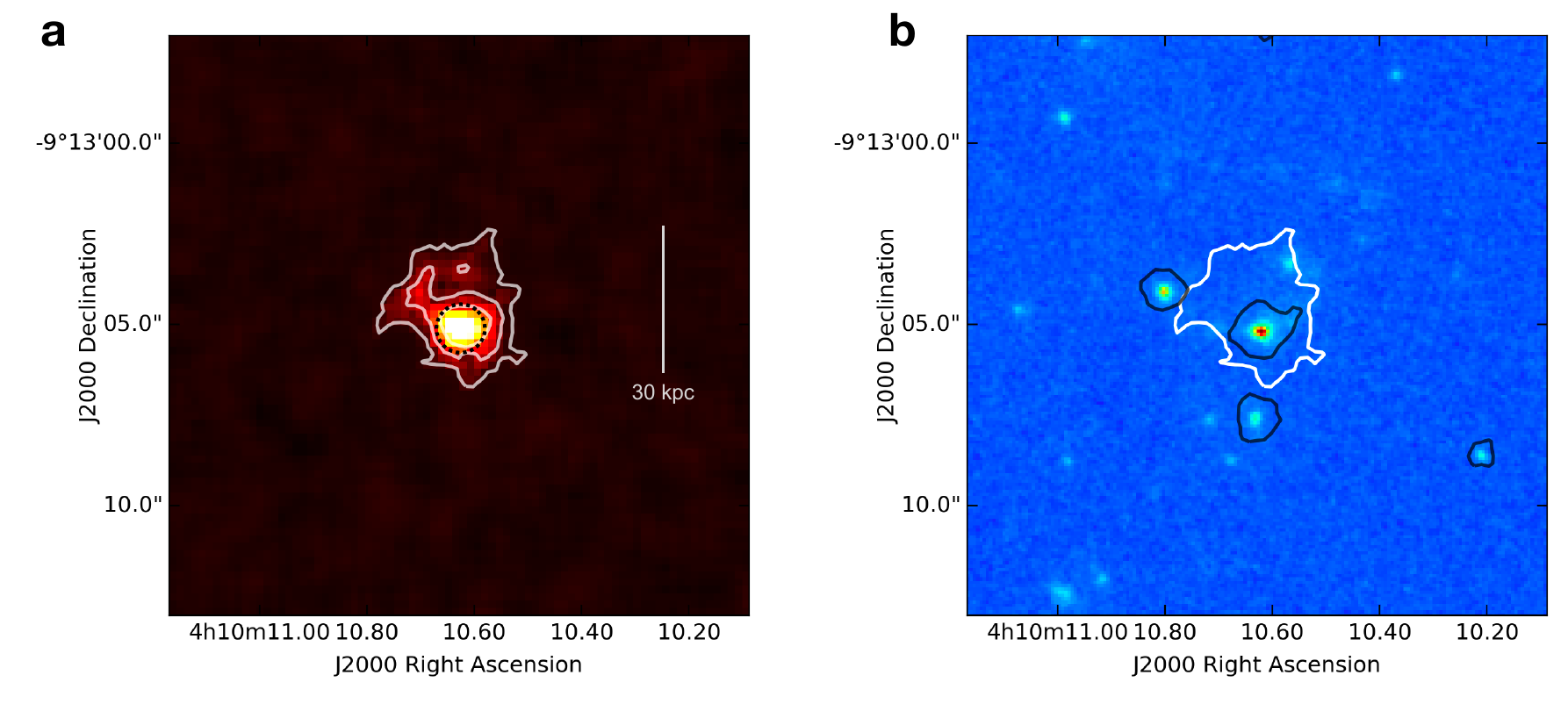}
	\caption{
		{\bf A small Lyman-$\alpha$ nebula, possibly associated to W0410-0913.}
		{\bf a} We show an optimally extracted map of the spatial distribution of the narrow and spectrally-offset Lyman-$\alpha$ emission emitted at the same location of the Hot DOG (see Figure \ref{fig:MUSEspectrum}).
		We measure a spatial extension, defined as the maximum projected linear size, of about $30 ~\rm kpc$ (see white line). 
		The white contours represent the $\rm [3, ~5, ~10]~\sigma$ levels of significance, 
		where the noise is estimated by propagating the variance and taking into account the number of layers contributing to each pixel in the 3D mask.
		The last contour at $3~\sigma$ defines the spatial projection of the 3D mask.
		For display purposes, we have added to the map of the projected 3D-mask one wavelength layer of the cube corresponding to the central wavelength of the nebula.
		{\bf b} We show the $3~\sigma$ contour of the nebula overlaid onto the HST WFC3 F160W image.
		Black contours are located around galaxies in the cutout that are detected in MUSE continuum (obtained collapsing the full datacube) and are defined as the regions in which $80\%$ of the flux is contained.  
	}
	\label{fig:MUSEnebula}
\end{figure*}

\begin{figure*}[ht]
	\centering
	\includegraphics[width=1\linewidth]{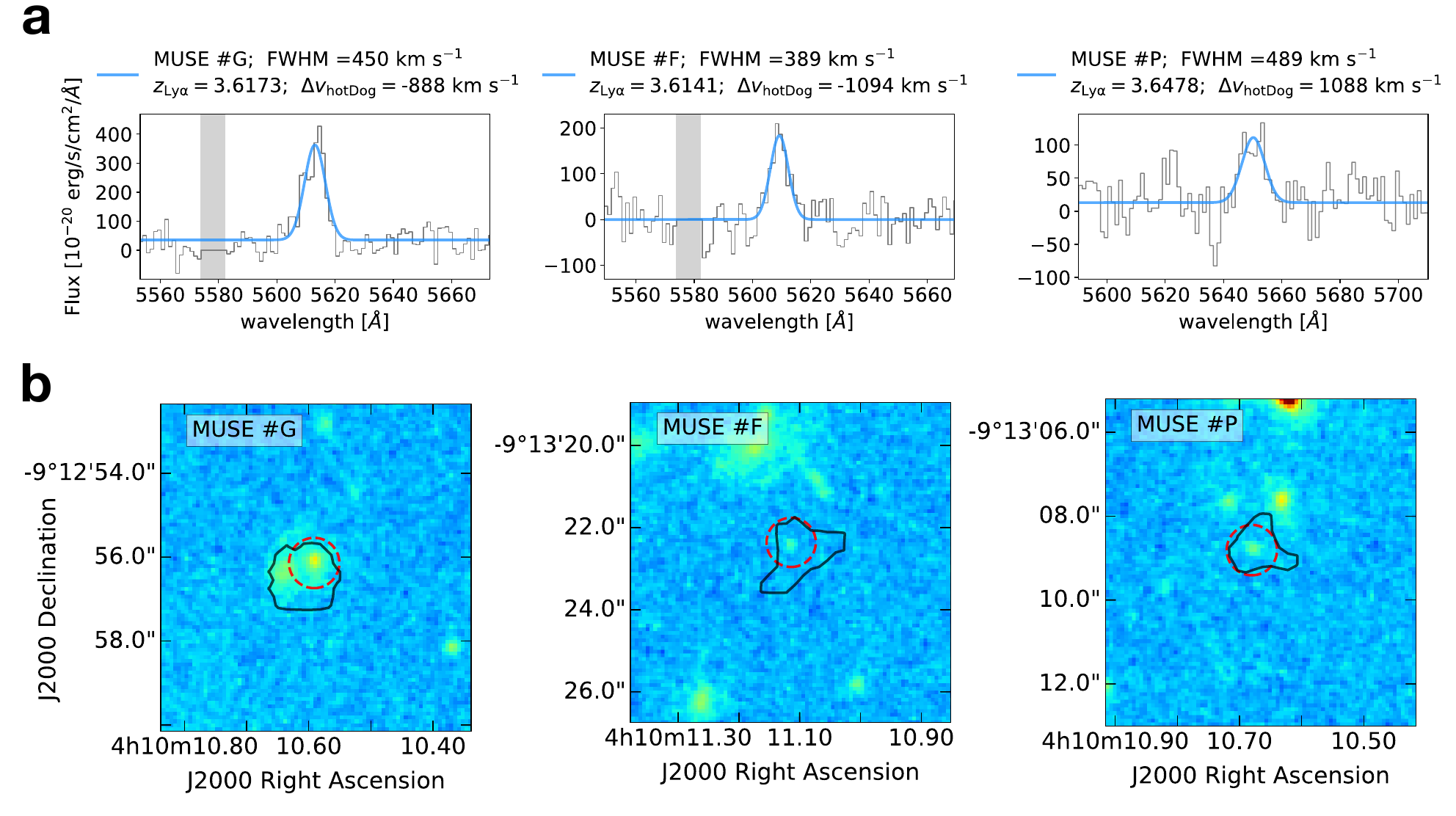}
	\caption{
		{\bf MUSE spectra and HST maps for a selection of companion galaxies around W0410-0913.}
		{\bf a} Lyman-$\alpha$ spectra of 3 among the Lyman-$\alpha$-emitters discovered with MUSE in the surrounding environment of W0410-0913. 
		These objects belong to the denser dynamic volume (with offset velocities in the range $[-2.500; +1.500] ~ \rm{km ~ s^{-1}}$), have HST detections, and are selected to be representative of the high (left), median (center) and low (right) distribution of the Lyman-$\alpha$ flux SNR (see Table 1).
		Lyman-$\alpha$ spectra, extracted from circular regions with a radius of $1''$ and centered around the Lyman-$\alpha$ peaks, are shown in gray. 
		The wavelength range in the x-axis has a width of $120 ~\AA$ and is centered around the central wavelength obtained through a Gaussian fit (Gaussian models are shown in blue). 
		Legends report the redshift of each source (derived through Lyman-$\alpha$), their FWHM, and their offset velocity with respect to the systemic redshift of the central luminous object. 
		Gray shaded areas represent the regions affected by sky lines.
		A compilation of MUSE spectra for all the Lyman-$\alpha$-emitting galaxies around W0410-0913 is shown in Supplementary Figures 1 and 2.
		{\bf b} HST cutouts ($4''\times4''$) of the Lyman-$\alpha$ shown above, detected with the WFC3/F160W filter (colored images).
		The $2\sigma$ contours of pseudo-NB Lyman-$\alpha$ images obtained with MUSE are overlaid in black.
		Red dashed circles represent the circular apertures with $1''$-sized radii that we use to extract the HST flux and estimate the SFR.
		The cutouts for all the HST-detected companion galaxies are shown in Supplementary Figure 3.
}
	\label{fig:selection_companions}
\end{figure*}

A collection of the optimally extracted images of the companion galaxies is shown in Figure \ref{fig:3}a.
We find that some of the luminous companion galaxies in the denser dynamic volume around W0410-0913, with offset velocities in the range $[-2.500; +1.500] ~ \rm{km ~ s^{-1}}$, show extended diffuse Lyman-$\alpha$ envelopes (not revealed, or barely detected, in the pseudo-NB images; see Figure \ref{fig:1}a) that are suggestive of connecting filamentary structures on scales up to $\gtrsim 100 ~{\rm kpc}$ (see Figure \ref{fig:3}).
We use the 3D-segmentation-masks provided by \texttt{CubEx} to produce also the 2D maps of the first moment (moment-1) of the flux distribution in its spectral domain. 
These maps give an indication of the centroid offset velocities with respect to the systemic redshift of W0410-0913 at each spatial location. 
A {\rm collage} of such kinematics maps is showed in Figure \ref{fig:3}b.

\subsection*{ALMA data}

In this work, we use archival ALMA observations of W0410-0913 targeting the emission-line spectra arising from rotational transitions of carbon monoxide CO(4-3), CO(6-5), and CO(7-6), and the atomic carbon transition [CI]$^3P_2-^3P_1$ (hereafter [CI]).
CO(4-3) $461 ~{\rm GHz}$, redshifted to $99.5 ~{\rm GHz}$ at $z=3.631$, was observed with ALMA in Band 3 on January 2018 (Program ID: 2017.1.00123.S), with a total integration time of $11.5~{\rm min}$. 
Deep observations of CO(6-5) $691.4 ~{\rm GHz}$, redshifted to $149.3 ~{\rm GHz}$, were taken in Band 4 in December 2017 (Program ID: 2017.1.00908.S) exploiting a total integration time of about $3.6~{\rm h}$.
CO(7-6) $806.6 ~{\rm GHz}$ and the adjacent [CI] $809.3 ~{\rm GHz}$, redshifted to $174.2 ~{\rm GHz}$ and $174.75 ~{\rm GHz}$ respectively, were observed in Band 5 on September 2018 (Program ID: 2017.1.00123.S), with a total integration time of about $10~{\rm min}$.
See the Data Availability Section for information on the data repositories.
We performed a standard calibration of the data using the Common Astronomy Software Applications package (\texttt{CASA} \cite{McMullin2007}), running different versions of the software per each dataset, according to the Quality Assurance report provided by the ALMA staff.
The pipeline calibration outputs did not show any issue, and thus we did not perform any additional flagging.
We obtained continuum-subtracted visibilities using the \texttt{CASA} task \texttt{uvcontsub}, and we generated the continuum-subtracted datacubes of the spectral windows containing the emission lines by running the \texttt{CASA} cleaning algorithm \texttt{clean} with 500 iterations and a SNR threshold of 3.
In order to maximize the sensitivity, we adopted a natural weighting of the visibilities for all the three datasets.
The resulting synthesized beam in the CO(4-3) datacube is $0.7'' \times 0.5''$ (position angle, ${\rm PA} = 83^\circ$), $0.22'' \times 0.18''$ (${\rm PA} = -69^\circ$) for CO(6-5), and $0.6'' \times 0.4''$ (${\rm PA} = 80^\circ$) in the datacube containing CO(7-6) and [CI].
We adopted pixel sizes of $0.1''$, $0.03''$ and $0.1''$, and spectral bins of $50~{\rm km~s^{-1}}$, $60~{\rm km~s^{-1}}$ and $50~{\rm km~s^{-1}}$ (which enable us to have both a good sampling of the lines and enough SNR per individual channel) for CO(4-3), CO(6-5) and CO(7-6) $+$ [CI] respectively.
The average sensitivities reached in the cubes (in spectral regions close to the line frequencies) are $0.7~ {\rm mJy~beam^{-1}}$ in a spectral channel of $50~{\rm km~s^{-1}}$ for CO(4-3), $0.08~ {\rm mJy~beam^{-1}}$ in a spectral channel of $60~{\rm km~s^{-1}}$ for CO(6-5) and $0.8~ {\rm mJy~beam^{-1}}$ in a spectral channel of $50~{\rm km~s^{-1}}$ for CO(7-6) $+$ [CI].

\begin{figure*}[ht]
	\centering
	\includegraphics[width=0.6\linewidth]{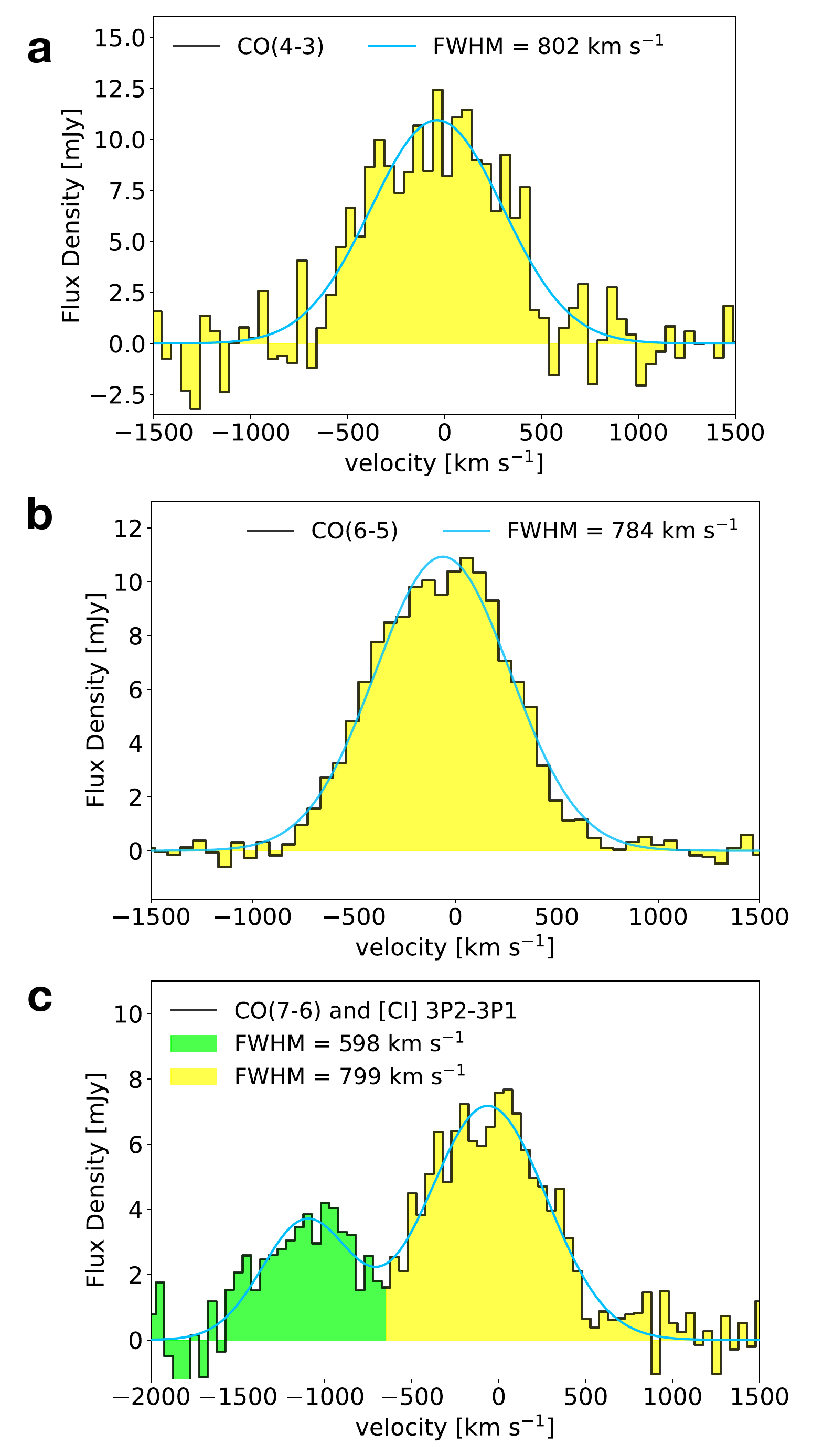}
	\caption{
		{\bf ALMA FIR lines observations of the ISM in W0410-0913.}		
		{\bf a} CO(4-3), {\bf b} CO(6-5), and {\bf c} CO(7-6) (yellow bars) and [CI] (green bars) spectra of W0410-0913 observed with ALMA.
		Gaussian models of the line profiles are shown with blue lines.
	}
	\label{fig:spectraALMA}
\end{figure*}

In Figure \ref{fig:spectraALMA} we show the CO(4-3), CO(6-5), CO(7-6) and [CI] spectra of W0410-0913.
These spectra are extracted from a spatial region defined by the connected pixels with ${\rm SNR} > 2$ in the  moment-0 maps, that we first obtain by collapsing the cube spectral channels in the range ${\rm [f_{cen} - FWHM: f_{cen} + FWHM]}$, where ${\rm f_{cen}}$ (the central frequency) and ${\rm FWHM}$ are derived by a Gaussian fit of a spectrum extracted from a $1''$-sized circular aperture centered at the expected position of the object.
Such a 2-iterations procedure helps in case of resolved observations to ensure that the whole line-emitting region is accounted for in the spectrum while avoiding signal dilution.
Through a non-linear least-squares Gaussian fit of the CO(6-5) line, which has the highest SNR among the available data, we derive a systemic redshift of $z_{\rm CO(6-5)} = 3.631 \pm 0.001$. 
This value is consistent with an independent measurement of $z_{\rm CO(6-5)}$ recently reported ~\cite{DiazSantos2021} ($z = 3.6301$), and the tiny difference is probably due to a slightly different definition of the extracting aperture.
The velocity axes of the spectra in  Figure \ref{fig:spectraALMA} are aligned to the common systemic redshift of  $z_{\rm CO(6-5)} = 3.631$.
From our Gaussian fit we derive large FWHMs of about $800~{\rm km~s^{-1}}$ (with average associated uncertainties of $70~{\rm km~s^{-1}}$) in CO(4-3), CO(6-5) and CO(7-6). [CI] appears to be narrower (${\rm FWHM} \sim 600~ \pm 90~{\rm km~s^{-1}}$), although this might be an effect of spectral blending with the more luminous CO(7-6).
Broad molecular line profiles in W0410-0913 were also reported in previously published ALMA observations of CO(4-3) ~\cite{Fan2018}, and [CII] \cite{DiazSantos2021}, although in both cases the instrument spectral setup was tuned on the optical redshift \cite{Wu2012}, which is blueshifted with respect to the systemic redshift as traced by the molecular gas, and the lines, close to the edge of the spectral window, were only partially detected.
Such broad emission lines in the integrated spectra extracted from the whole galaxy reflect the rapidly spinning nature of its disk (see subsection Kinematic modeling of the CO(6-5) emission in the ISM of W0410-0913).

\subsection*{Study of the CGM environment with ALMA}

Unfortunately, none of these datasets, designed to observe the cold gas emission from the hyper-luminous W0410-0913, provides information about the rest-frame FIR line properties of the companion Lyman-$\alpha$ emitters that we discover with MUSE in its CGM.
However this is not surprising, and it is mainly due to a combination of insufficient sensitivity and too high resolution (that dilutes the flux over several beams), as we test by running simulations of the estimated telescope time on the ALMA Observing Tool (OT; version released for Cycle 8).
In detail, we first estimate the total molecular gas mass ($M_{\rm mol}$) from the SFR measured with HST by using standard $M_{\rm mol}$-SFR scaling relations calibrated at high-$z$ \cite{Tacconi2018}.
We then convert $M_{\rm mol}$ into total CO fluxes by adopting a conservative Milky-Way-like conversion factor between $M_{\rm mol}$ and the CO(1-0) luminosity, $\alpha^{\rm MW}_{\rm CO} = 4.3$, and standard Milky-Way-like intensity ratios between CO(1-0) and the rotational transitions observed by ALMA \cite{Carilli2013}.
We note that the adopted value of $\alpha_{\rm CO}$ has been calibrated at solar metallicities, and should be considered conservative in our case since $\alpha_{\rm CO}$ typically increases at lower metallicities \cite{Hunt2020}, and high-$z$ galaxies tend to be less metal-rich at a fixed stellar mass \cite{Maiolino2019}.
Assuming projected sizes of about $1''$ and line-widths of $300 ~{\rm km~s^{-1}}$, we find that the estimated total ALMA time to observe the integrated CO spectrum in a galaxy with ${\rm SFR} = 30 ~{\rm M_{\odot}~yr^{-1}}$ (the median SFR of the HST-detected companions),  with at least $3\sigma$ per channel in at least 3 spectral bins, at the same angular resolution of the available observations, would be about $1.6 ~{\rm h}$ for CO(4-3) and CO(7-6), and about $24.5 ~{\rm h}$ for CO(6-5), between a factor of $6-8$ longer than the observed datasets.
Other additional limitations derive from the ALMA smaller fields of view (the primary beams range from about $30''$ in Band 5 to about $45''$ in Band 3), and the narrower spectral coverage (sidebands ranging from  $5000 ~{\rm km~s^{-1}}$ in Band 5 to  $9000 ~{\rm km~s^{-1}}$ in Band 3) with respect to MUSE data.
The combination of these effects reduces the number of companion galaxies that could be possibly detected (up to $60~\%$, in the best case) and, together with the sensitivity issue discussed above, limits the information that can be derived through stacking analyses.
Indeed we find signals that are consistent with noise in the stacked images that we produce, for each dataset, by combining moment-0 maps centered at the expected positions of the galaxies (using the peak of Lyman-$\alpha$ emission as a prior), obtained by collapsing three different velocity ranges, $300, ~400, ~500 ~{\rm km~s^{-1}}$, centered around the expected systemic frequencies (using the redshifts derived from the Lyman-$\alpha$ as a prior).

\subsection*{Kinematic modeling of the CO(6-5) emission in the ISM of W0410-0913}

The deep and highly resolved CO(6-5) observations of W0410-0913 are best suited to study the molecular gas kinematics in W0410-0913, that we model using the 3D tilted ring model fitting code $^{\rm{3D}}$Barolo \cite{DiTeodoro2015} (3D-Based Analysis of Rotating Objects from Line Observations).
This technique has the advantage that it directly models the 3D cube taking beam-smearing effects into account and it has been extensively applied to observations of high-redshift ($z>2$) galaxies \cite{Jones2021}.
The fit routine is performed on the continuum-subtracted CO(6-5) datacube and can be summarized as follows.

First, we isolate the signal by producing a 3D signal mask of the emission line through the $^{\rm{3D}}$Barolo SEARCH tool (based on DUCHAMP \cite{Whiting2012}). 
In detail, this algorithm identifies pixels in the datacube with intensities above a certain SNR (SNR$_{\rm upper}$ = 3.5, in our case), and then it searches around these signal peaks for emission above a second SNR threshold (i.e., SNR$_{\rm lower}$ = 2.0, in our case).
The $^{\rm{3D}}$Barolo best-fit results are highly dependent on the overall geometry of the model (i.e., number and width of the rings) and on the initial estimates of the parameters.
Thus, once the signal is isolated we create the moment maps needed for the initial guesses. 
Using the CASA toolkit task {\tt image.moments} we collapse the cube over the spectral channels containing the line and produce a moment-0 map, which we then fit with a 2D Gaussian (using the CASA task {\tt image.fitcomponents}) to obtain the FWHM of the major and minor axes, the position angle and the central position. 
We also produce a velocity dispersion map (i.e., moment-2) including only the pixels that belong to the 3D mask identified before. 
Then, we fit a tilted ring model to the line emission by using the $^{\rm{3D}}$Barolo key function {\tt 3DFIT}. 
We adopt a value of twice the FWHM of the major axis of the restoring beam as a maximum model radius. 
As the value for the width of each ring, we adopt the FWHM of the minor axis, in order to not over-resolve the data\cite{Jones2021}.
This corresponds to a number of 12 model rings.
The model central position is fixed to be coincident with the centroid of a 2D Gaussian fit of the moment-0 map, and we use the maximum value in the moment-2 map as the initial guess of the velocity dispersion.
A thin disk, with a scale height of 0.01$''$, is assumed.
We adopt an initial guess for the inclination of 45$^\circ$, and we allow it to vary between 10-80$^\circ$.
Starting from this set of guesses, $^{\rm{3D}}$Barolo produces a model of the first ring by randomly populating the volume with discrete emitting gas clouds in order to reproduce the position angle, the rotational velocity, etc of the initial estimates.
The resulting model datacube is then convolved with the ALMA synthesized beam and rescaled so that the velocity-integrated intensity of each spaxel in the data and model cubes are equal.
Then, the values of the pixels in the data and model datacubes are compared, and each parameter is varied until the absolute residual (i.e., |model - observation|) is minimized. 
When this process reaches a minimum, the next ring is analyzed. 
During the first run {\tt 3DFIT} provides best-fit results for the rotational velocity, velocity dispersion, systemic velocity, inclination, and position angle. The three latter variables are averaged across all the rings, and the whole fit routine is run again, with only the rotational velocity and velocity dispersion as variables.
Among the final $^{\rm{3D}}$Barolo outputs, we obtain morphological parameters (e.g., central position), and kinematic parameters such as inclination, position angle, velocity dispersion, and rotation.

\begin{figure*}[ht]
	\centering
	\includegraphics[width=1\linewidth]{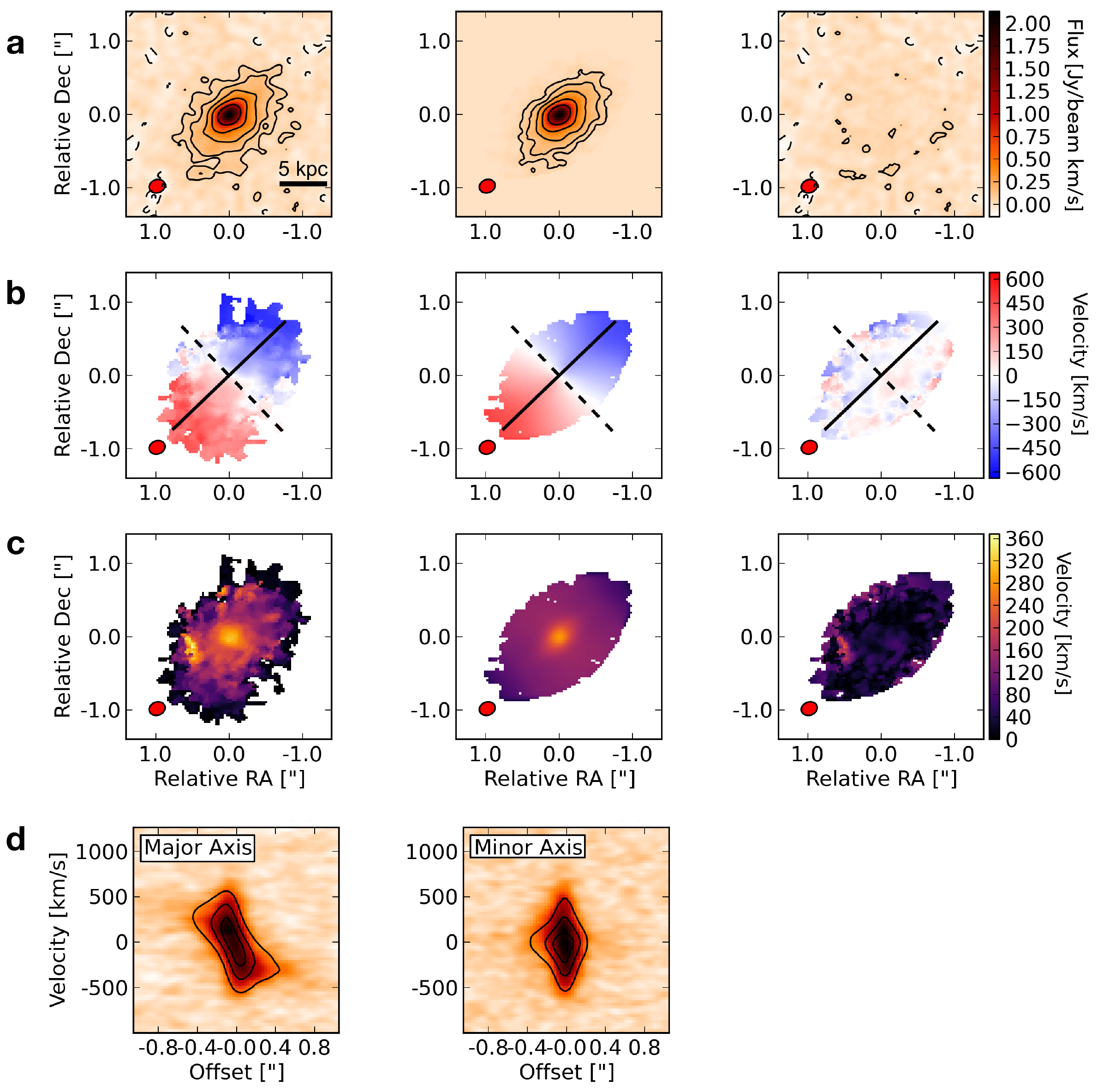}
	\caption{
		{\bf 
				Moment maps of ALMA CO(6-5) observations compared with the $^{\rm{3D}}$Barolo models.
		}
		In panel {\bf a}, {\bf b} and {\bf c},
		we show the moment-0, moment-1,
		and moment 2 of the CO(6-5) data. 
		For each of these panels, the three columns denote (from left to right) the observed
		data, models (created with $^{\rm{3D}}$Barolo), and the corresponding residuals.
		The black solid line in the lower-left corner of the moment-0 panel shows a 5 kpc physical scale.
		The solid lines in the second row represent the kinematic major axis, while the dashed lines represent the minor axis.
		The black contours in the moment-0 maps are defined as in Figure \ref{fig:4}.
		In panel {\bf d} we show the major axis (left) and the minor axis (right) PVDs. 
		For each PVD, the data is shown by the background color, while the overlaid contours
		represent the model at 20\%, 50\%, and 80\% of its maximum value.
		The ALMA beam, colored in red, is shown in the lower-left corners.
	}
	\label{fig:barolo}
\end{figure*}

In  Figure \ref{fig:barolo} we show that our model is able to reproduce the kinematic properties of the target to an excellent level, with small residuals in the moment maps and a good agreement between the model and data position-velocity diagram (PVDs).
The galaxy has a well-defined rotation, with rotational velocities that range between $v_{\rm rot} \rm \sim \pm ~500 ~ km ~ s^{-1}$, as can be seen from the moment-1 map, showing the spatial distribution of line-of-sight velocities, and from the PVDs extracted along the major and minor axes (see Figure \ref{fig:barolo}). 
Despite the fast coherent rotation, the velocity dispersion is still large, in analogy with other recent ALMA observations of Hot DOGs reporting turbulent ISM \cite{DiazSantos2021}.
Nevertheless, as discussed in the Article, we find that $v_{\rm rot} / \sigma > 1$ at each ring in our model (see Figure \ref{fig:4}).
We find morphological and kinematic position angles of $133 \pm 5^\circ$ and $133.8 \pm 4.9^\circ$ respectively, and morphological and kinematic inclination on the plane of the sky of $49.5 \pm 3.3^\circ$ and $50.1 \pm 13.1^\circ$.

\section*{Data Availability}

The MUSE data used in this work (programs: 0102.A-0729, 106.2184-001) are available at \url{http://archive.eso.org/scienceportal/home}.
The archival HST (program ID: 12930) and ALMA (programs: \#2017.1.00123.S,
\#2017.1.00908.S) data are available at \url{https://archive.stsci.edu/}
and \url{http://almascience.eso.org/aq/} respectively.
All data generated and/or analyzed during the current study are available from the corresponding author upon request.

\section*{Code Availability}

The MUSE pipeline recipes, used to reduce the MUSE data, are publicly available and the software distribution can be obtained at \url{https://www.eso.org/sci/software/pipelines/muse/}.
The software \texttt{CASA}, used to process the ALMA data, is publicly available and can be downloaded at \url{https://almascience.eso.org/processing/science-pipeline}.
The code used to model the molecular gas kinematics, $^{\rm{3D}}$Barolo, is publicly accessible through github (\url{https://editeodoro.github.io/Bbarolo/}), with a full documentation hosted at \url{https://bbarolo.readthedocs.io/en/latest/}.
The package used for the 3D extraction of faint diffuse emission in the MUSE datacube, \texttt{CubExtractor}, is not publicly available, though the reader interested in using this code can contact Sebastiano Cantalupo at the mail address \url{sebastiano.cantalupo@unimib.it}.

\bibliographystyle{naturemag-doi}
\bibliography{bib.bib}

\section*{Acknowledgments}

ALMA is a partnership of ESO (representing its member states), NSF (USA) and NINS (Japan), together with NRC (Canada), NSC and ASIAA (Taiwan), and KASI (Republic of Korea), in cooperation with the Republic of Chile. The Joint ALMA Observatory is operated by ESO, AUI/NRAO and NAOJ.
The CosmoSim database used in this paper is a service by the Leibniz-Institute for Astrophysics Potsdam (AIP).
The Cosmic Dawn Center (DAWN) is funded by the Danish National Research Foundation under grant No. 140.
G.C.J. acknowledges funding from European Research Council (ERC) Advanced Grants 695671 ``QUENCH’’ and 789056 ``FirstGalaxies’’ under the European Union’s Horizon 2020 research and innovation programme, as well as support by the Science and Technology Facilities Council (STFC).
F.R. acknowledges support from the European Union’s Horizon 2020 research and innovation program under the Marie Sklodowska-Curie grant agreement No. 847523 ‘INTERACTIONS’. 
S.CAN gratefully acknowledges support from the ERC under the European Union’s Horizon 2020 research and innovation programme grant agreement No 864361.
R.M. acknowledges support by the Science and Technology Facilities Council (STFC), by the ERC Advanced Grant 695671 “QUENCH” and also funding from a research professorship from the Royal Society.
E.P. acknowledges support from PRIN MIUR project "Black Hole winds and the Baryon Life Cycle of Galaxies: the stone-guest at the galaxy evolution supper", contract \#2017PH3WAT. L.G. and R.S. acknowledge support from the Amaldi Research Center funded by the MIUR program ”Dipartimento di Eccellenza” (CUP:B81I18001170001).

\section*{Author information}

M.G., E.P., L.Z. and R.M. designed the MUSE observing strategy and wrote the MUSE proposals, with the help of F.A.B., S. Cantalupo, S. Carniani, L.P., G.C.J., A.T., R.V., L.G. and R.S.
~ M.G. analyzed the MUSE data, with the help of A.T.
~ M.G., G.C.J., S.Carniani, and F.R. analyzed the ALMA data.
~ M.G. and L.P. analyzed the HST data.
~ C.D.B., K.K., P.L., V.M., F.S., and A.V. contributed to the interpretation of the MUSE data.
~ N.M. led the calculation on the physical significance of the overdensity. 
All authors contributed to the interpretation and the discussion of the scientific results.
~ M.G. wrote the manuscripts and all authors provided comments.
\newline
Correspondence and requests for materials should be addressed to M.G. (email: michele.ginolfi@eso.org).

\section*{Competing Interests}
The authors declare no competing interests.

\section*{Tables}

\begin{table*}\label{tab:table}
	\centering
	\begin{tabular}{llllllll}
		\hline
		~  & ~ & ~ &~ & ~ & ~ & ~ & ~\\ [-1.5ex]
		ID  & RA & DEC  & $z_{\rm Ly\alpha}$  & ${\rm FWHM}_{\rm Ly\alpha}$  & $L_{\rm Ly\alpha}$ & SNR & ${\rm SFR}$  \\  [+0.5ex]
		~  &[deg]  & [deg] & ~  & [km s$^{-1}$]& ${\rm log}(L_{\rm Ly\alpha} / {\rm L_\odot})$  & ~  & [${\rm M_\odot ~ yr^{-1}}$] \\  [+0.5ex]
		\hline 
		~  & ~&  ~& ~ & ~ & ~ & ~\\ [-1.5ex]
		$\#A$ & 	62.5433 &  $-$9.21704		 &		 3.5425 $\pm$ 0.0003		& 		250 $\pm$ 52 		&		 42.21 $\pm$ 0.28 	& 7.1  &    21 $\pm$ 3		 \\ [+0.5ex]
		$\#B$ & 	62.5442 &  $-$9.21871		 &		 3.5947 $\pm$ 0.0001		& 		215 $\pm$ 15 		&		 42.38	$\pm$ 0.41 	 &  17.7 &57 $\pm$ 3  \\ [+0.5ex]
		$\#C$ & 	62.543 &  $-$9.22171		 &		 3.6100 $\pm$ 0.0003		& 		292 $\pm$ 50		&		 42.02	$\pm$ 0.32 	& 6.1  &	$--$ \\ [+0.5ex]
		$\#D$ & 	62.5437 &  $-$9.22193		 &		 3.6103 $\pm$ 0.0003		& 		382 $\pm$ 62		&		 42.15	$\pm$ 0.33 &  6.9 &	$--$  \\ [+0.5ex]
		$\#E$ & 	62.5418 &  $-$9.22037		 &		 3.6133 $\pm$ 0.0003		& 		374 $\pm$ 66		&		 42.11	$\pm$ 0.31	&  7.0 &	$--$  \\ [+0.5ex]
		$\#F$ & 	62.5463 &  $-$9.22299		 &		 3.6141 $\pm$ 0.0003		& 		389 $\pm$ 50		&		 42.24 $\pm$ 0.39  & 8.9  & 12 $\pm$ 3		 \\ [+0.5ex]
		$\#G$ & 	62.544 &  $-$9.2156		 &		 3.6173 $\pm$ 0.0002		& 		450 $\pm$ 32		&		 42.55	$\pm$ 0.51 	   & 16.6  & 64 $\pm$ 3	 \\ [+0.5ex]
		$\#H$ & 	62.5433 &  $-$9.21782		 &		 3.6181 $\pm$ 0.0004		& 		367 $\pm$ 75		&		 41.98 $\pm$ 0.28  & 5.6  &	$--$  \\ [+0.5ex]
		$\#I$ & 	62.5411 &  $-$9.22176		 &		 3.6236 $\pm$ 0.0002		& 		280 $\pm$ 34 		&		 42.13	$\pm$ 0.39 	 & 9.7  & $--$  \\ [+0.5ex]
		$\#J$ & 	62.5432 &  $-$9.2196		 &		 3.6252 $\pm$ 0.0002		& 		232 $\pm$ 41		&		 41.95 $\pm$ 0.31 	 & 6.4 &  $--$ \\ [+0.5ex]
		$\#K$ & 	62.5434 &  $-$9.21904		 &		 3.6255 $\pm$ 0.0002		& 		195 $\pm$ 32		&		 41.94 $\pm$ 0.29 	&  6.4 &	$--$	 \\ [+0.5ex]
		$\#L$ & 	62.542 &  $-$9.21932		 &		 3.6256 $\pm$ 0.0002		& 		127 $\pm$ 29		&		 41.81 $\pm$ 0.25 		&  4.9 &	$--$ \\ [+0.5ex]
		$\#M$ & 	62.5513 &  $-$9.21365		 &		 3.6360 $\pm$ 0.0009		& 		701 $\pm$ 170		&		 42.15 $\pm$ 0.25 		&  4.5 &$--$	 \\ [+0.5ex]
		$\#N$ & 	62.5436 &  $-$9.21437		 &		 3.6426 $\pm$ 0.0004		& 		498 $\pm$ 69		&		 42.20	$\pm$ 0.35 	& 7.7  &$--$	 \\ [+0.5ex]
		$\#O$ & 	62.5456 &  $-$9.21504		 &		 3.6466 $\pm$ 0.0004		& 		528 $\pm$ 69		&		 42.16 $\pm$ 0.35 	& 7.5 &	$--$	 \\ [+0.5ex]
		$\#P$ & 	62.544 &  $-$9.21937		 &		 3.6478 $\pm$ 0.0005 		& 		489 $\pm$ 99		&		 42.08	 $\pm$ 0.20 & 5.6  &	23 $\pm$ 3	 \\ [+0.5ex]
		$\#Q$ & 	62.5456 &  $-$9.21382		 &		 3.6495 $\pm$ 0.0002		& 		555 $\pm$ 45		&		 42.48 $\pm$ 0.47 	& 14.0  &	$--$	 \\ [+0.5ex]
		$\#R$ & 	62.5432 &  $-$9.21843		 &		 3.6504 $\pm$ 0.0005		& 		647 $\pm$ 90 		&		 42.32 $\pm$ 0.36 	&8.1  &	$--$	 \\ [+0.5ex]
		$\#S$ & 	62.5464 &  $-$9.21743		 &		 3.6514 $\pm$ 0.0003		& 		242 $\pm$ 54 		&		 41.89$\pm$ 0.27 	&  5.5&	$--$	 \\ [+0.5ex]
		$\#T$ & 	62.539 &  $-$9.21532		 &		 3.6760 $\pm$ 0.0005		& 		399 $\pm$ 90		&		 41.92	$\pm$ 0.24 	& 4.8   & 40 $\pm$ 3	 \\ [+0.5ex]
		$\#U$ &  62.5486 & $-$9.21315		 &		 3.7305 $\pm$ 0.0002		& 		451 $\pm$ 40		&		 42.44	$\pm$ 0.46  	& 13.0  &	16 $\pm$ 3 \\ [+0.5ex]
		$\#V$ & 	62.5446& $-$9.22293		 &		 3.7326 $\pm$ 0.0001		& 		420 $\pm$ 25		&		 42.66$\pm$ 0.55 	&  18.9 &	34 $\pm$ 3	 \\ [+0.5ex]
		$\#W$ & 	62.5372& $-$9.22532		 &		 3.7328 $\pm$ 0.0003		& 		454 $\pm$ 55		&		 42.36 $\pm$ 0.4 	& 9.8  &	33 $\pm$ 3	 \\ [+0.5ex]
		$\#X$ & 	62.5449& $-$9.21782		 &		 3.7880 $\pm$ 0.0002		& 		327 $\pm$ 41		&		 42.16	$\pm$ 0.39 	& 8.5  & 99 $\pm$ 3	 \\ [+0.5ex]
		\hline
		~  & ~&  ~& ~& ~&~& ~&\\
	\end{tabular}
	\caption{
		{\bf Measured properties of the Lyman-$\alpha$ emitters in the overdensity.}
		In the first, second and third columns we report the IDs and the coordinates (Right Ascension and Declination) associated to the companion galaxies detected through Lyman-$\alpha$ in the surroundings of W0410-0913.
		The coordinates correspond to the centroid of Lyman-$\alpha$ emission in the pseudo-NB images.
		Redshifts, FWHMs and $L_{\rm Ly\alpha}$ (n the fourth, fifth and sixth column, respectively) are derived from the MUSE spectra as described in the Methods, and their errors represent the uncertainties provided by the Gaussian models.
		The SNR values in the seventh column are computed as the ratio between the integrated flux in the MUSE spectra 
		and the noise associated to the width of the line (for each source, we considered spectral regions as large as $2\times$ FWHM).
		In the last column, the SFRs and their uncertainties are based on the HST images and are computed as discussed in the subsection Measure of the SFR from HST in Methods.
	}
\end{table*}

\section*{Figures}

\section*{Supplementary Information}

\begin{Supplementary Figure*}[ht]
	\centering
	\includegraphics[width=1\linewidth]{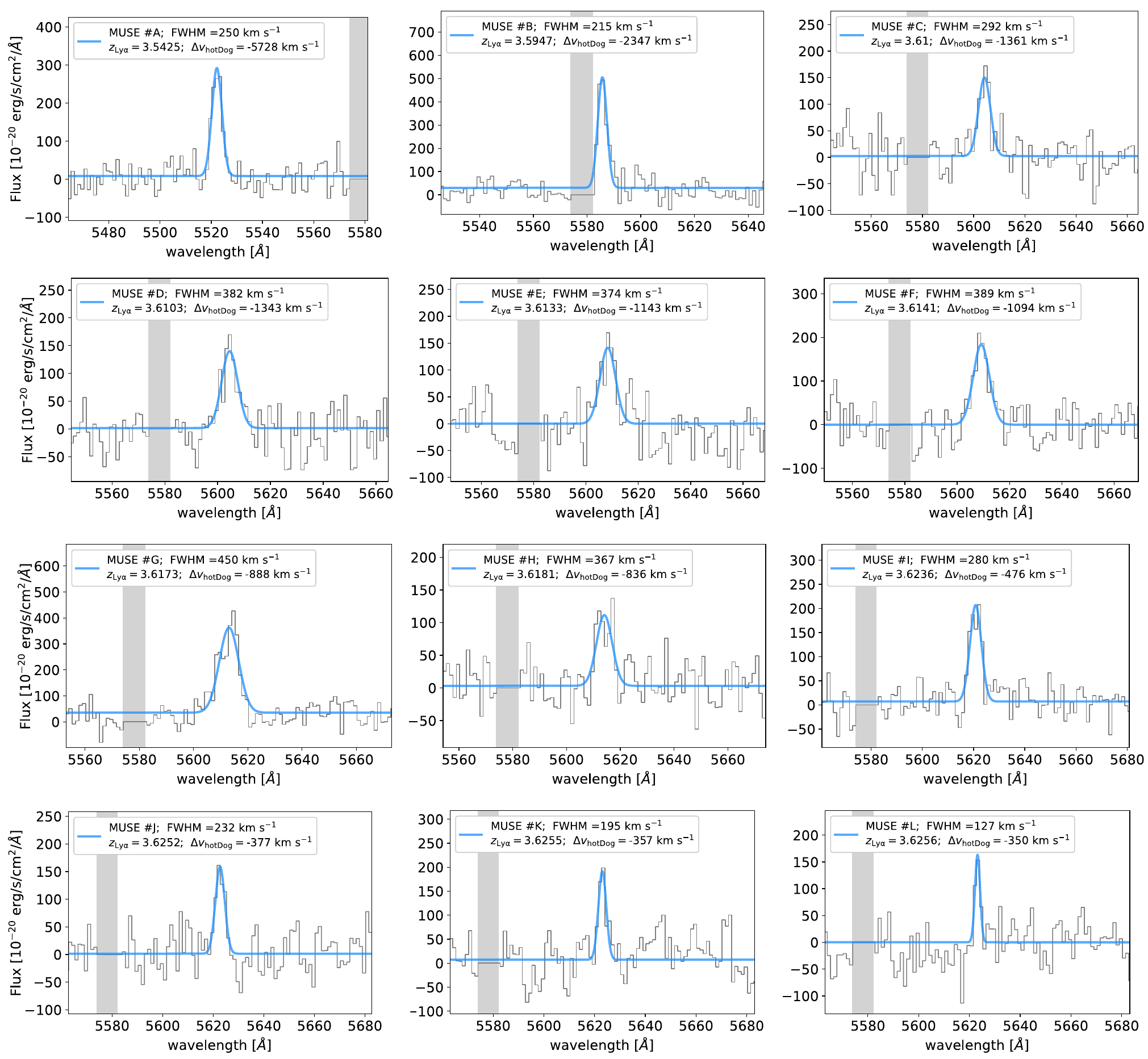}
	\caption{
		{\bf Lyman-$\alpha$ spectra of the companion galaxies discovered with MUSE in the surrounding environment of W0410-0913.}
		Lyman-$\alpha$ spectra, extracted from circular regions with a radius of $1''$ and centered around the Lyman-$\alpha$ peaks, are shown in gray. 
		The wavelength range in the x-axis has a width of $120 ~\AA$ and is centered around the central wavelength obtained through a Gaussian fit (Gaussian models are shown in blue). 
		Legends report the redshift of each source (derived through Lyman-$\alpha$), their FWHM, and their offset velocity with respect to the systemic redshift of the central luminous object.
		Gray shaded areas represent the regions affected by sky lines. 		
		See Table 1 for the definition of IDs and a summary of the measured properties of the Lyman-$\alpha$ emitters.
		Figure continues in Supplementary Figure 2.
	}
\end{Supplementary Figure*}

\begin{Supplementary Figure*}[ht]
	\centering
	\includegraphics[width=1\linewidth]{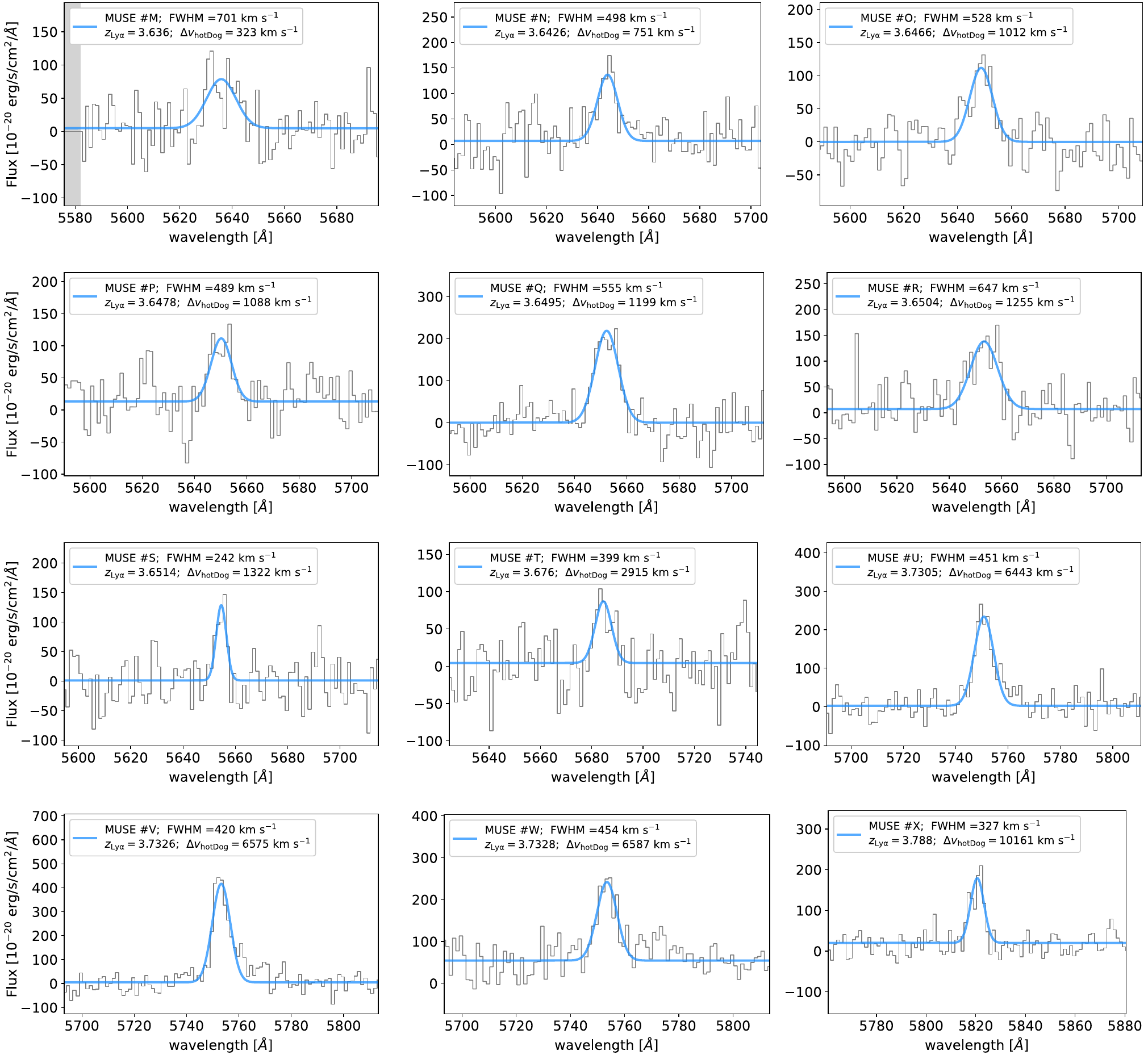}
	\caption{
		{\bf Lyman-$\alpha$ spectra of the companion galaxies discovered with MUSE in the surrounding environment of W0410-0913.}
		Lyman-$\alpha$ spectra, extracted from circular regions with a radius of $1''$ and centered around the Lyman-$\alpha$ peaks, are shown in gray. 
		The wavelength range in the x-axis has a width of $120 ~\AA$ and is centered around the central wavelength obtained through a Gaussian fit (Gaussian models are shown in blue). 
		Legends report the redshift of each source (derived through Lyman-$\alpha$), their FWHM, and their offset velocity with respect to the systemic redshift of the central central luminous object.
		Gray shaded areas represent the regions affected by sky lines.
		See Table 1 for the definition of IDs and a summary of the measured properties of the Lyman-$\alpha$ emitters.
		Figure starts in Supplementary Figure 1.
	}
\end{Supplementary Figure*}

\begin{Supplementary Figure*}[ht]
	\centering
	\includegraphics[width=0.8\linewidth]{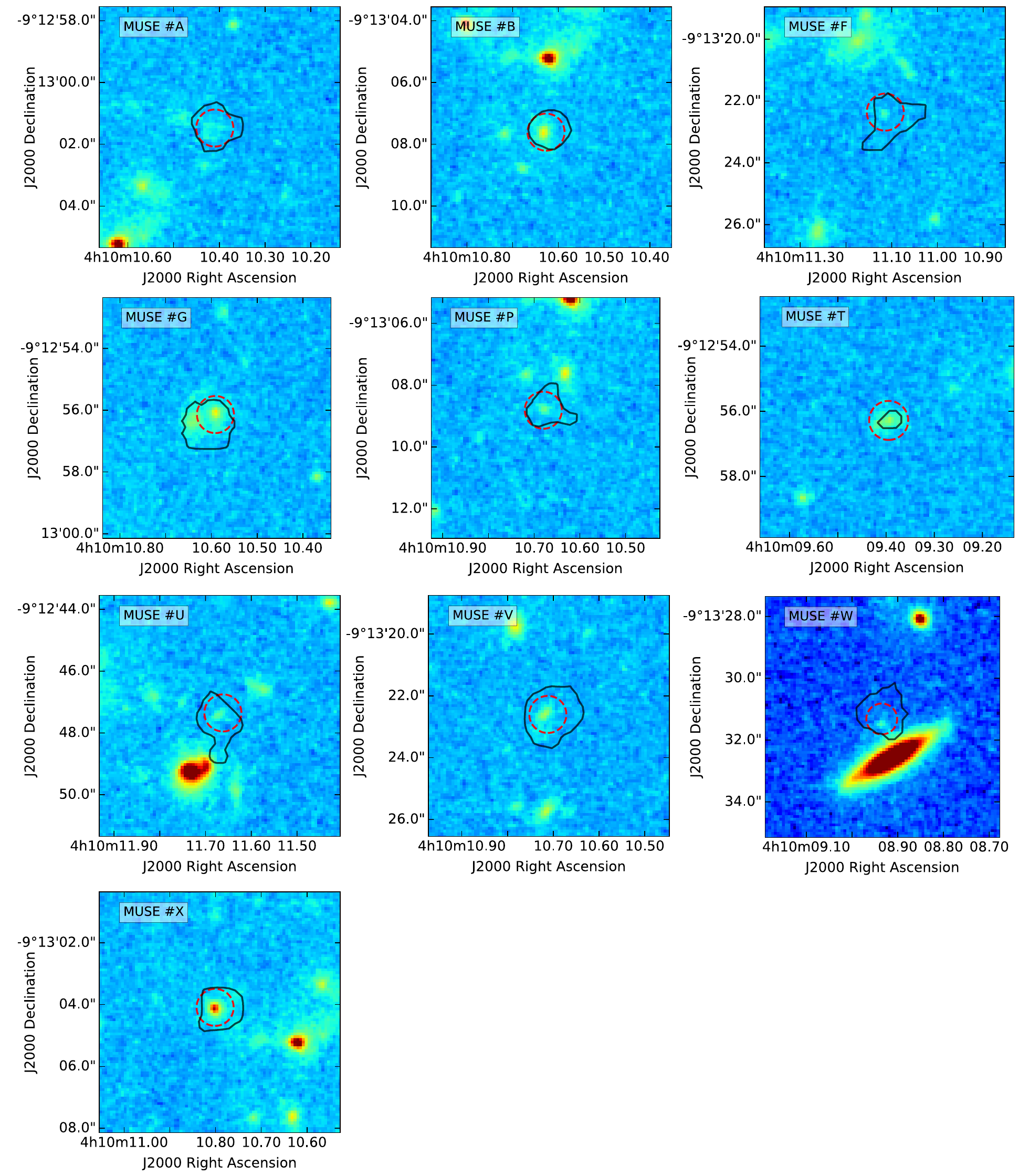}
	\caption{
		{\bf Maps of the HST-detected companion galaxies.}
		The HST cutouts ($4''\times4''$) of the Lyman-$\alpha$ emitters detected with the WFC3/F160W filter are shown (colored images).
		The $2\sigma$ contours of pseudo-NB Lyman-$\alpha$ images obtained with MUSE are overlaid in black.
		Red dashed circles represent the circular apertures with $1''$-sized radii that we use to extract the HST flux and estimate the SFR.
	}
	\label{fig:HST_muse}
\end{Supplementary Figure*}


\end{document}